\documentclass[a4paper,11pt]{article}
\pdfoutput=1 % if your are submitting a pdflatex (i.e. if you have
\usepackage{graphicx}
\usepackage{float}
\usepackage{mciteplus}
\usepackage{dirtytalk}

\usepackage{jheppub} % for details on the use of the package, please
\usepackage[T1]{fontenc} % if needed

\renewcommand{\[}{\begin{equation}}
\renewcommand{\]}{\end{equation}}
\def\beq{\begin{equation}}
\def\eeq{\end{equation}}
\newcommand{\be}{\begin{eqnarray}}
\newcommand{\ee}{\end{eqnarray}}

\renewcommand{\texttt}{{}}

\def\bs{\begin{subequations}}
\def\es{\end{subequations}}

\def\Bc{\mathcal{B}}
\def\Cc{\mathcal{C}}
\def\Dc{\mathcal{D}}

\def\Fc{\mathcal{F}}

\def\Hc{\mathcal{H}}

\def\Lc{\mathcal{L}}
\def\Mc{\mathcal{M}}
\def\Nc{\mathcal{N}}
\def\Oc{\mathcal{O}}
\def\Pc{\mathcal{P}}

\def\Rc{\mathcal{R}}

\def\Tc{\mathcal{T}}

\newcommand{\tia}[1]{}

\newcommand{\bea}{\begin{eqnarray}}
\newcommand{\eea}{\end{eqnarray}}
\newcommand{\beas}{\begin{eqnarray*}}
\newcommand{\eeas}{\end{eqnarray*}}
\newcommand{\bal}{\begin{aligned}}
\newcommand{\eal}{\end{aligned}}

\def\({\left(}
\def\){\right)}

\newcommand{\LF}{\left(}
\newcommand{\RF}{\right)}
\newcommand{\LT}{\left[}
\newcommand{\RT}{\right]}

\newcommand{\pd}{\partial}

\newcommand{\const}{\mathrm{const}}

\title{Non-Gaussianities in generalized non-local $R^2$-like inflation}

\author[a,b,f]{Alexey S. Koshelev,}

\author[c,d]{K. Sravan Kumar,}

\author[e,f]{Alexei A. Starobinsky}

\affiliation[a~]{
	School of Physical Science and Technology, ShanghaiTech University, 201210 Shanghai, China
}
\affiliation[b~]{
	Departamento de F\'isica, Centro de Matem\'atica e Aplica\c{c}oes (CMA-UBI),
	Universidade da Beira Interior, 6200 Covilh\~a, Portugal }
\affiliation[c~]{Department of Physics, Tokyo Institute of Technology, 2-12-1 Ookayama, Meguro-ku, Tokyo 152-8551, Japan}
\affiliation[d~]{ Institute of Cosmology \& Gravitation,
	University of Portsmouth,
	Dennis Sciama Building, Burnaby Road,
	Portsmouth, PO1 3FX, United Kingdom}
\affiliation[e~]{
	L. D. Landau Institute for Theoretical Physics RAS, Chernogolovka, Moscow region 142432,
	Russian Federation}
\affiliation[f~]{Kazan Federal University, Kazan 420008, Republic of Tatarstan, Russian
	Federation}

\emailAdd{askoshelev@shanghaitech.edu.cn}
\emailAdd{sravan.kumar@port.ac.uk}
\emailAdd{alstar@landau.ac.ru}

\abstract{	In \cite{Koshelev:2022olc}, a most general higher curvature non-local gravity action was derived that admits a particular $R^2$-like inflationary solution predicting the spectral index of primordial scalar perturbations $n_s(N)\approx 1-\frac{2}{N}$, where $N$ is the number of e-folds before the end of inflation, $N\gg 1$, any value of the tensor-to-scalar ratio $r(N)<0.036$ and the tensor tilt $n_t(N)$ violating the $r= -8n_t$ condition. In this paper, we compute scalar primordial non-Gaussianities (PNGs) in this theory and effectively demonstrate that higher curvature non-local terms lead to reduced bispectrum $f_{\rm NL}\LF k_1,\,k_2,\,k_3 \RF$ mimicking several classes of scalar field models of inflation known in the literature. We obtain $\vert f_{\rm NL}\vert \sim O(1-10)$ in the equilateral, orthogonal, and squeezed limits and the running of these PNGs measured by the quantity $\vert\frac{d\ln f_{\rm NL}}{d\ln k}\vert\lesssim 1$. Such PNGs are sufficiently large to be measurable by future CMB and Large Scale Structure observations, thus providing a possibility to probe the nature of quantum gravity. Furthermore, we demonstrate that the $R^2$-like inflation in non-local modification of gravity brings non-trivial predictions which go beyond the current status of effective field theories (EFTs) of single field, quasi-single field and multiple field inflation. A distinguishable feature of non-local $R^2$-like inflation compared to local EFTs is that we can have running of PNGs at least an order of magnitude higher. In summary, through our generalized non-local $R^2$-like inflation, we obtain a robust geometric framework of inflation that can explain any detection of observable quantities related to scalar PNGs. }

\makeatother

\begin{document}
	
\maketitle

\section{Introduction}

Present CMB observations strongly support the $R^2$ (or Starobinsky) inflation where quasi-de Sitter (dS) expansion is a result of a phase when Ricci scalar $R$ being an eigenmode of d'Alembertian operator \cite{Starobinsky:1980te}, see also \cite{Starobinsky:1981vz,Starobinsky:1983zz,Vilenkin:1985md,Mijic:1986iv,Maeda:1987xf}. In a recent paper of the authors \cite{Koshelev:2022olc}, a most general higher curvature non-local gravity action that leads to $R^2$-like inflationary phase with a scalaron and a massless graviton degrees of freedom was obtained. It is motivated from several quantum gravity frameworks, especially the proposition that an action of quantum gravity should be an extension of general relativity (GR) with all possible curvature invariant terms one can write. The generalized non-local $R^2$-like inflation developed in \cite{Koshelev:2022olc} was shown to be compatible with the observed scalar spectral index $n_s = 0.9649\pm 0.0042$ at $68\%~{\rm CL}$ \cite{Akrami:2018odb} and any value of the tensor-to-scalar ratio consistent with the latest upper bound $r<0.036$ \cite{BICEP:2021xfz}. It also admits both positive and negative values of the tensor spectral index $n_t$ and violation of the single field tensor consistency relation $r= -8n_t$.  Furthermore, it was found that the  generalized non-local $R^2$-like inflation \cite{Koshelev:2022olc} does not fall into any categories of so-called effective field theory of  single field inflation (EFT-SI) \cite{Cheung:2007st,Weinberg:2008hq} since the framework of gravity is non-local in nature and does not introduce any non-trivial sound speeds for the perturbed degrees of freedom. In this paper, we compute 3-point scalar correlations or the scalar Primordial Non-Gaussianities (PNGs) of generalized non-local $R^2$-like inflation and confront our results against the various limits of  the reduced bispectrum $f_{\rm NL}$ \cite{Chen:2010xka,Kenton:2015lxa} that are constrained from the latest CMB data as follows \cite{Planck:2019kim}\footnote{In the literature \say{squeezed} {limit} $f_{\rm NL}^{\rm sq}$ is most often called as \say{local} {limit} \cite{Chen:2010xka,Planck:2019kim}, but we stick to the definition of \say{squeezed} {limit} \cite{Kenton:2015lxa} just in order not to confuse with the meaning of \say{locality} in this paper.}
\begin{equation}
	f_{\rm NL}^{\rm sq}= -0.9\pm 5.1,\quad  f_{\rm NL}^{\rm equiv} = -26\pm 47,\quad f^{ortho}_{\rm NL} = -38\pm 24 \,\,{\rm at}\,\, 68\,\%~{\rm CL}\,.
	\label{boundsfnl}
\end{equation}
PNGs in the case of standard single field (canonical scalar field) inflation are very small to be detectable (i.e., $f_{\rm NL}\sim O(10^{-2})$), but models based on non-canonical scalar(s) (also generalized scalar-tensor theories like Horndeski \cite{Kobayashi:2011nu}) and constructions with non adiabatic vacuum (non-Bunch-Davies) initial conditions are known to give large PNGs with respect to equilateral ($f^{\rm eq}_{\rm NL}$) and orthogonal ($f^{\rm orth}_{\rm NL}$) {limits}, whereas squeezed {limit} ($f^{\rm sq}_{\rm NL}$) is understood to be the feature of multiple scalar fields or non-slow-roll inflation or non-trivial initial conditions for quantum states \cite{komatsu2009nongaussianity,Chen:2010xka,Bahrami:2013isa}. There has been plethora of inflationary models developed in the context of effective field theories (EFTs) to address different {types of} PNGs \cite{Akrami:2019izv,Cheung:2007st,Chen:2006nt,Chen:2005fe,Creminelli:2010qf,Senatore:2010wk}. 
Furthermore, recent developments of quasi-single field inflation present new signatures for PNGs due to the presence of heavy fields coupled to the inflaton \cite{Arkani-Hamed:2015bza,Chen:2009zp}. 
In summary, any detection of PNGs, i.e., $f_{\rm NL}\sim \Oc(1-10)$ with the present constraints \cite{Akrami:2019izv} is majorly understood to be the only feature of physics beyond the standard canonical scalar field inflation that are mentioned above. However, in this paper a new understanding of the PNGs emerges in the context of analytic non-local gravity  where the nature of propagating degrees of freedom significantly deviate from the standard effective field theories (EFTs) of inflation in which ``locality" is preserved, but the dispersion relations are modified due to time dependent parameters multiplying the spatial derivative terms. 

Our framework of study is how higher curvature non-local gravity affects $f_{\rm NL}\LF k_1,\,k_2,\,k_3 \RF$ {which is a function of three momenta}. We ultimately obtain a clear conclusion that {several signatures of PNGs of $f_{\rm NL}\sim O(1)$ can be well explained} by geometrical modifications of gravity mimicking several classes of scalar-tensor theories. We also study the scale dependent nature of bi-spectrum in our theory of inflation. We envisage all the possible observational features of our geometric non-local $R^2$-like inflation in the context of 3-point scalar correlations and provide a detailed discussion to distinguish our framework with large class of EFTs of inflation involving scalar fields. 

The paper is organized as follows. 
In \textbf{Sec.~\ref{sec:results}} we provide a concise summary of results obtained in this paper on PNGs in generalized non-local $R^2$-like inflation. In \textbf{Sec.~\ref{sec:sum}} we review briefly the generalized non-local gravity action obtained in \cite{Koshelev:2022olc}. 
In \textbf{Sec.~\ref{sec:NGs}} and \textbf{Sec.~\ref{sec:NGRUN}} we compute PNGs of generalized non-local $R^2$-like inflation and discuss several {limits} of PNGs which arise from different dimensionless free-parameter values. We compute the running of PNGs defined by the Non-Gaussianity spectral index $n_{\rm NG} = \frac{d\ln f_{\rm NL}}{d\ln k }$ and provide details of reconstructing the formfactors in the theory through future cosmological observations with the ultimate aim to probe the scale of non-locality. We quantitatively elucidate in detail the scale dependent nature of reduced bispectrum in this theory. 
In \textbf{Sec.~\ref{sec:eftcom}} we provide an extensive discussion of distinguishing the features of generalized non-local $R^2$-like inflation against several class of EFTs of inflation involving single and multiple scalar fields. We show how generalized non-local $R^2$-like inflation {enhances} our understanding of early Universe cosmology. 
In \textbf{Sec.~\ref{sec:Conc}} we discuss further studies which are possible to perform beyond the scope of this paper. 
In \textbf{Appendix~\ref{App3NG}} we provide additional details for the computation of PNGs performed in Sec.~\ref{sec:NGs} and Sec.~\ref{sec:NGRUN}. In \textbf{Appendix~\ref{app:violmc}} we provide a heuristic explanation for violation of the Maldacena consistency relation in non-local gravity despite having slow-roll and single field behaviour during inflation and the adiabatic vacuum initial conditions.

\textbf{Notations:} In this paper, our metric signature is $\LF -,\,+,\,+,\,+ \RF$. We use overdot and $^\prime$ denotes derivative with respect to cosmic time (t) and conformal time ($ \tau $) respectively, $^\dagger$,\,$^{\dagger\dagger}$,\,$^{\dagger\dagger\dagger}$ for first, second and third derivative with respect to the argument. 
We use overbar to denote background quantities for flat Friedmann-Lema\^itre-Robertson-Walker (FLRW) spacetime, 4-dimensional indices are labelled by small Greek letters and three dimensional quantities are denoted by $i,j =1,2,3$. We also set $\hbar=c=1$ and $M_p=1/\sqrt{8\pi G}$ is the reduced Planck mass. Everywhere we perform the computations in the leading order in slow-roll parameters ($\epsilon\sim \frac{1}{N}$ approximation).
%with $N$ being the number of e-folds before the end of inflation. 
The subscript \say{$_{\rm dS}$} denotes the quantities in the quasi-dS approximation. 

\section{Summary of our results}
\label{sec:results}

Primordial Non-Gaussianities (PNGs) are an important probe to understand the nature of self-interactions of primordial fields during inflation. In this paper, we studied the scalar PNGs emerging from generalized non-local $R^2$-like inflation which is developed in \cite{Koshelev:2022olc} from the quest of finding a most general quantum gravity action compatible with the so far observed physics of early Universe cosmology.  
Our results show that  it is possible to generate interesting {limits} of scalar PNGs in a geometric framework of inflation (i.e, extension of GR with higher derivative and higher curvature terms). The novelty of our study is that we obtain several interesting PNGs by just changing the dimensionless finite number of free parameters of the  generalized non-local $R^2$-like inflation \cite{Koshelev:2022olc}. Most importantly, the predictions of PNGs we report in this paper do not affect the observables of scalar and tensor power spectrum such as 
\begin{equation}
	\Biggl\{ n_s,\,\frac{dn_s}{d\ln k},\,r,\, n_t,\, \frac{dn_t}{d\ln k},\,\frac{d^2n_t}{d^2\ln k} \Biggr\}
\end{equation}
which is often the case in the several class of inflationary models based on the scalar fields \cite{Martin:2013tda} (see Sec.~\ref{sec:eftcom} for an extensive discussion). The generalized non-local gravity action \cite{Koshelev:2022olc} given in \eqref{NAID} describes the most general gravity theory that leads to $R^2$-like inflation. The action  \eqref{NAID} contains analytic infinite derivative (non-local) terms quadratic and cubic in scalar curvature $R$ and non-local terms involving Weyl tensor square and Weyl tensor cube etc. First, we prove that in the leading order of the slow-roll approximation (but accounting for all corrections from infinite derivative terms), the scalar PNGs in this theory are generated only from the terms that do not contain Weyl tensor. The terms involving Weyl tensor are relevant for tensor PNGs which are the subject of future investigation \cite{KKS2}. Secondly, we compute scalar PNGs in the theory and discover various {limits} which appear as a result of changing the (finite) parameter space of the theory. Most importantly, we obtain new {class of} PNGs {with scale dependent runnings} which are due to the non-local terms cubic in scalar curvature that is a new revelation in comparison with the earlier constructions of non-local $R^2$-like inflation with quadratic curvature terms\footnote{The results we report here are slightly different from \cite{Koshelev:2020foq} since our choice of formfactors is different from the one considered in  \cite{Koshelev:2020foq} as explained in \cite{Koshelev:2022olc}. } \cite{Koshelev:2020foq}. Furthermore, we compute running of PNGs in this theory which turned out to be most interesting aspect of non-local theory in the scope of future cosmological observations. 

The central message of our study is that large PNGs\footnote{By large PNGs we literally mean $f_{\rm NL}\sim O(1-10)$, that is ten to hundred times higher than the prediction of standard canonical scalar field inflation \cite{Chen:2010xka}. } can appear in a purely geometric modification of gravity in a very non-trivial way breaking all the known theorems of EFT of inflation which states \cite{Chen:2010xka,komatsu2009nongaussianity} that large scalar PNGs can only be generated if one goes beyond the following conditions:
\begin{itemize}
	\item canonical single field inflation with the speed of sound $c_s=1$;
	\item standard slow-roll; 
	\item adiabatic vacuum initial conditions (often called the Bunch-Davies vacuum).
\end{itemize}
In our case, we satisfy all the above criteria but still we generate {various} PNGs mimicking large class of EFT models of scalar field inflation (see Sec.~\ref{sec:eftcom} for more details). This result of ours is completely due to non-local interactions of curvature perturbations, and therefore we {extend further} the so far established notions of PNGs and the primordial physics \cite{komatsu2009nongaussianity,Planck:2019kim,Planck:2015zfm}. Below we enlist different {class of} PNGs we obtained in this model which mimic the {PNGs} of several scalar field EFT of inflation (for each case we report here the maximum values of $f_{\rm NL}$ we can obtain with full detailed analysis done in Sec.~\ref{sec:NGs}). But 
contrary to the conventional models of inflation, $f_{\rm NL}$ in our case gets a strong scale dependence (see explanation around \eqref{fnlseo} to \eqref{Rfnlseo}) 
which allow us to distinguish our gravity theory from EFTs which we discussed in  Sec.~\ref{sec:eftcom}. 

\begin{itemize}
	\item Mimicking non-canonical single field inflation $f^{\rm eq}_{\rm NL}\sim f^{\rm orth}_{\rm NL} \sim O(10)$ and $f^{\rm sq}_{\rm NL}\sim O\LF 10^{-2} \RF$ \cite{Chen:2006nt}. This possibility is obtained for the case of parameter space explained in Fig.~\ref{R2zsq}.  PNG peaked in the equilateral template is usually recognized as a feature of a non-canonical scalar field inflation (e.g., so called DBI inflation) where the primordial curvature perturbation propagates with a non-trivial sound speed ($c_s$) \cite{Chen:2006nt}. Here we obtain {this PNG} due to non-local interactions of curvature perturbation whose sound speed is Unity. 
	\item Mimicking non-canonical multifield inflation with $f_{\rm NL}\sim O(1-10)$ \cite{Gao:2008dt}. We obtain this PNG in the case of non-local gravity action \eqref{NAID} with and without the non-local cubic terms in Ricci scalar (see Fig.~\ref{fig:fnlqcg}, Fig.~\ref{R2CFNL} and Fig.~\ref{R2zeq}). In the case without non-local cubic terms in Ricci scalar, i.e., $\lambda_c\to 0$ in  \eqref{NAID}, we can obtain $f_{\rm NL}\sim O(1)$ and a robust  relation between different {limits} of $f_{\rm NL}$ stated in \eqref{relQG} which is independent of the formfactor. Furthermore,   in  Fig.~\ref{R2CFNL} we depict {PNGs} with $f^{\rm eq}_{\rm NL}\sim f^{\rm orth}_{\rm NL} \sim O(10)$ and $f^{\rm sq}_{\rm NL}\sim 1$. Notably, we violate the well-known Maldacena consistency relation for the squeezed limit $f_{\rm}^{\rm sq} =\frac{5}{12}\LF 1-n_s \RF$ despite having single-field slow-roll regime. This is truly due to non-local interactions of quantum fluctuations that represents a new physics in the context of PNGs (we explained this effect in detail with a simple example in Appendix~\ref{app:violmc}.) 
	\item Mimicking multifield field inflation and/or inflation with non-adabatic initial conditions $f^{\rm sq}_{\rm NL}\sim f^{\rm orth}_{\rm NL} \sim 1$ and $f^{\rm eq}_{\rm NL}\sim O\LF 10^{-2} \RF$ \cite{Byrnes:2014pja,Chen:2010xka,Bahrami:2013isa}. 
	This curious type of PNG is obtained for the parameter space of the theory presented through Fig.~\ref{R2zsor}. 
\end{itemize}
All the above {class of various PNGs} are obtained by changing various dimensionless parameters of the theory \eqref{NAID}. We obtain various types of PNGs in a single theory with non-local purely geometric modification of gravity. This is the novel finding of this work which enforces the need to {enhance} our understanding of early Universe cosmology. We also computed running of various {limits} of PNGs in this model in \eqref{Rfnlseo} and deduced that $\Bigg\vert\frac{d\ln f_{\rm NL}}{d\ln k} \Bigg\vert \lesssim 1$ that can be one to two orders of magnitude higher than in conventional models of inflation (see Sec.~\ref{sec:eftcom} for details). Running PNGs could play a pivotal role in inflationary cosmology and they could be potentially detected in future CMB probes \cite{Chen:2005fe,Oppizzi:2017nfy}. 

All the above predictions make the generalized non-local $R^2$-like inflation a viable target for future CMB and Large Scale Structure observations aimed to detect $f_{\rm NL} \sim O(1)$  \cite{Meerburg:2019qqi,Karagiannis:2018jdt,Castorina:2018zfk,Munoz:2015eqa,Floss:2022grj}. We argue that just the detection of  $f_{\rm NL} \sim O(1)$ does not confirm the nature of inflaton. As this work elucidates, one can generate all different PNGs within the framework of a non-local higher curvature modification of gravity. Therefore, we stress that future observations must focus on the running of PNGs in order to probe the primordial physics correctly. 

\begin{figure}
	\centering
	\includegraphics[width=0.6\linewidth]{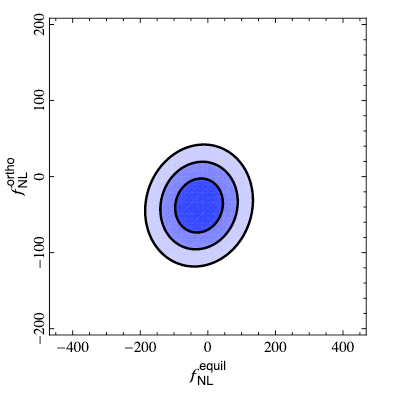}
	\caption{68\,\%, 95\,\%, and 99.7\,\% confidence regions $\LF f_{\rm NL}^{\rm eq},\,f_{\rm NL}^{\rm orth} \RF$ taken from \cite{Planck:2019kim}. The predictions of generalized non-local $R^2$-like inflation lie well within the bounds in this plot along with the squeezed limit PNG $-6<f_{\rm NL}^{\rm sq}< 4.2$. We obtain $\LF f_{\rm NL}^{\rm eq},\,f_{\rm NL}^{\rm orth} \RF \sim O(10)$, and various {limits} of $f_{\rm NL}$ are presented in detail in Sec.~\ref{sec:NGs} and Sec.~\ref{sec:NGRUN}.}
	\label{fig:pngpl}
\end{figure}

\section{Generalized non-local $R^2$-like inflation}
\label{sec:sum}
We briefly present here the details of generalized non-local $R^2$-like inflation established in detail in \cite{Koshelev:2022olc}.  The $R^2$ inflationary background is actually the spatially flat FLRW  solution of the following eigen-value equation 
\begin{equation}
	\square R=M^2 R 
	\label{mR2in}
\end{equation}
where $ R$ is the Ricci scalar, $\square$ is the d'Alembertian operator and $M^2$ is the mass of the scalaron \cite{Starobinsky:1980te}. The recent Planck observations of the scalar spectral index $n_s$ strongly support \eqref{mR2in} (that corresponds to the so-called exponential-like Plateau inflation in the Einstein frame). Motivated by this fact, in \cite{Koshelev:2022olc} 
a most general non-local gravity action that is compatible with the $R^2$ inflationary background \eqref{mR2in} has been constructed which is  
\begin{equation}
	\begin{aligned}
		S_H^{\rm Non-local} = & \int d^4x\sqrt{-g} L_H^{\rm Non-local} \\
		= 		 &\int d^4x\sqrt{-g} \Bigg(\frac{M_{p}^{2}}{2}R +
		\frac{1}{2}\Bigg[ R\Fc_R\LF \square_s \RF R   +  \LF \frac{M_p^2}{2\Mc_s^2}+ f_0 R_s \RF W_{\mu\nu\rho\sigma}{\mathcal{F}}_{W}\left(\square_{s},\, R_s \right)W^{\mu\nu\rho\sigma}  \\
		&+ \frac{f_0\lambda_c}{\Mc_s^2}\Lc_1\LF \square_{s} \RF R\, \Lc_2\LF \square_{s} \RF R\, \Lc_3\LF  \square_{s}\RF R \\
		& +\frac{f_0\lambda_R}{\Mc_s^2}\Dc_1\LF \square_s \RF R\Dc_2\LF \square_s\RF W^{\mu\nu\gamma\lambda}\Dc_3\LF \square_s \RF W_{\mu\nu\gamma\lambda}
		\\
		& +\frac{f_0\lambda_W}{\Mc_s^2}\Cc_1\LF \square_s \RF W_{\mu\nu\rho\sigma}\Cc_2\LF \square_s\RF W^{\mu\nu\gamma\lambda}\Cc_3\LF \square_s \RF W_{\gamma\lambda}^{\quad\rho\sigma}\Bigg]+\cdots\Bigg)\, ,
		\label{NAID}
	\end{aligned}
\end{equation}
where $M_p$ is the reduced Planck mass, $R$ is the Ricci scalar, $W_{\mu\nu\rho\sigma}$ is the Weyl tensor, $f_0 =\frac{M_p^2}{6M^2} $ with $M$ is being the scalaron mass, $\square_s = \frac{\square}{\Mc_s^2}$ with $\Mc_s$ being the so-called scale of non-locality, $R_s = \frac{R}{\Mc_s^2}$ and
\begin{equation}
	\Bigg\{ \Fc_R\LF \square_s \RF,\,\Fc_W\LF \square_s,\, \frac{2R}{3\Mc_s^2}\RF,\,\Lc_i\LF \square_s \RF,\,\Cc_i\LF \square_s \RF,\,\Dc_i\LF \square_s \RF\Bigg\}
	\label{formfactl}
\end{equation} 
are the formfactors which are the analytic non-polynomial functions. In \cite{Koshelev:2022olc} it was explicitly shown that these operators should take the following form in order avoid ghost degrees of freedom in the Minkowski and quasi-de Sitter (dS) limits: 
\begin{equation}
	\begin{aligned}
		\Fc_R\LF \square_s \RF & = f_0 M^2 \frac{1-\LF 1-\frac{\square}{M^2}  \RF e^{\gamma_S\LF \square_s \RF} }{\square_s}\, ,\\
		\Fc_W\LF  \square_s,\,\frac{R}{M_s^2} \RF &  = 
		\frac{e^{\gamma_T
				\LF \square_s-\frac{2}{3}R_s \RF}-1}{\square_s-\frac{2}{3}R_s}
	\end{aligned}
	\label{formfinal}
\end{equation}
where $\gamma_S\LF \square_s \RF$ and $\gamma_T\LF \square_s-\frac{2}{3}\frac{R}{\Mc_s^2} \RF$ are the entire functions which can be polynomials or functions of polynomials \cite{Koshelev:2022olc}.  
The formfactors 
$\Lc_i\LF \square_s \RF,\, \Cc\LF \square_s \RF,\, \Dc_i\LF \square_s \RF$ should be such that they are suppressed at the large momentum limit $p\to \infty$. 
These formfactors can in general contain an infinite number of arbitrary parameters, but here we assume them to be defined by a finite number of unknown parameters. Our assumption here is motivated from an expectation that any UV-complete theory can be formulated with a finite number of free parameters which can be probed by some observations.  In Sec.~\ref{sec:NGs} we deduce that the terms involving Weyl tensor do not contribute to the scalar PNGs which we are interested in. Therefore, the details of the operators $\Dc_i\LF \square_s \RF,\, \Cc_i\LF \square_s \RF$ are irrelevant to the present study. 
To illustrate our results of scalar PNGs, we express the operators $\Lc_i\LF \square_s \RF$ in the following form.
\begin{equation}
	\begin{aligned}
		\Lc_i\LF \square_s \RF =  e^{\ell_i\LF\square_s\RF}-1\, ,
	\end{aligned}
	\label{cubicformd}
\end{equation}
where $\ell_{i}\LF \square_s \RF$ are the entire functions which we assume to be polynomials or functions of polynomials that gives us a finite parameter space. 
  In order to have \eqref{mR2in} as a particular exact FLRW background solution of the theory \eqref{NAID}, the following conditions are sufficient \cite{Koshelev:2022olc}
\begin{equation}
	\gamma_S\LF \frac{M^2}{\Mc_s^2} \RF =0,\quad \ell_{i}\LF \frac{M^2}{\Mc_s^2} \RF =0\, ,
	\label{entic}
\end{equation}
which we call as on-shell conditions. 
Since Weyl tensor is zero on FLRW backgrounds equations of motion are trivially satisfied and we do not need any conditions on $\gamma_T\LF \square_s-\frac{2}{3}\frac{R}{\Mc_s^2} \RF$, $\Cc_i\LF \square_s \RF$ and $\Dc_i\LF \square_s \RF$ at the background level. From \eqref{entic} we can write a generic form of entire functions as 
\begin{equation}
	\begin{aligned}
		\gamma_S\LF \square_s \RF = & \LF \square_s-\frac{M^2}{\Mc_s^2} \RF P_S\LF \square_s \RF \, , \\ 
		\ell_{i}\LF \square \RF = & \LF \square_s-\frac{M^2}{\Mc_s^2} \RF G_i\LF \square_s \RF  
	\end{aligned}
	\label{entchoice}
\end{equation}
where $P_S\LF \square_s \RF,\, G_i\LF \square_s \RF$ are the finite degree polynomials such that the operators $\Lc_i\LF \frac{p^2}{\Mc_s^2} \RF$ do not grow in the limit $p\to \infty$, for the theory to be consistent in the UV-regime and to apply it safely to the low-energy context which is inflation in our case. As a  result, the parameter space of the theory \eqref{NAID} is finite dimensional although the action \eqref{NAID} is non-local (i.e., infinite derivative) in nature. It was shown in \cite{Koshelev:2022olc} that the cubic non-local scalar curvature term does not contribute to the second order perturbed action but this term does contribute to the third order perturbed action that will be explained in the next section. 
Furthermore, we would like to emphasize here the PNGs we compute and report in the later sections are fully compatible with the predictions of scalar and tensor power spectrum and their tilts reported in \cite{Koshelev:2022olc}. Finally, note that the terms denoted by $\cdots$ in \eqref{NAID} are the higher order non-local scalar curvature and Weyl curvature terms which we assume to be not relevant for the 3-point inflationary correlations (see the discussion around \eqref{r4nl}). 

\section{PNGs and their running in generalized non-local $R^2$-like inflation}

\label{sec:NGs}

This section is dedicated to computing scalar PNGs of $R^2$-like inflation in our higher curvature non-local action \eqref{NAID}. We make use of several calculations in this regard from the previous study  \cite{Koshelev:2020foq} where the computation of PNGs in non-local theories was robustly developed. 
Let us start with recalling some standard procedure and definitions related to the computation of 3-point correlations which are defined by \cite{Maldacena:2002vr,Koshelev:2020foq} 
\begin{equation}
	\langle\Rc\left(\mathbf{k_{1}}\right)\Rc\left(\mathbf{k_{2}}\right)\Rc\left(\mathbf{k_{3}}\right)\rangle  = -i \int_{-\infty}^{\tau_e}   d\tau \langle 0\vert [ \Rc(\tau_{e},\,\mathbf{k_1})\Rc(\tau_{e},\,\mathbf{k_2})\Rc(\tau_{e},\,\mathbf{k_3}),\, H_{int} ] \vert 0 \rangle \,,
	\label{3-point-f}
\end{equation}
where $\boldsymbol{k}_i$ are wave vectors and $H_{int}\approx -\Lc_3$ is the interaction Hamiltonian that is approximately equal to the 3rd order perturbation of the Lagrangian (\ref{NAID}) ($\Lc_3$) within the slow-roll approximation  \cite{Maldacena:2002vr,DeFelice:2011zh} and $\tau_e$ denotes the end of inflation.

The so-called bi-spectrum ($\Bc_\Rc$) is defined as
\begin{eqnarray}
	\langle\Rc\left(\mathbf{k_{1}}\right)\Rc\left(\mathbf{k_{2}}\right)\Rc\left(\mathbf{k_{3}}\right)\rangle  =  \left(2\pi\right)^{3}\delta^{3}\left(\mathbf{k_{1}}+\mathbf{k_{2}}+\mathbf{k_{3}}\right)\mathcal{B}_{\Rc}\left(k_{1},k_{2},k_{3}\right)
	\label{3corr}
\end{eqnarray}
where $\vert\boldsymbol{k_i}\vert=k_i$ and 
the non-linear curvature perturbation $\Rc$ is expressed as 
\cite{Komatsu:2001rj,Takahashi:2014bxa}
\begin{equation}
	\Rc = \Rc_g  -\frac{3}{5}f_{NL}\LF \Rc_g^2-\langle \Rc_g \rangle ^2 \RF \,, 
\end{equation} 
where $\Rc_g$ being the Gaussian random field and the $f_{NL}$ is the non-linearity parameter also known as the reduced bi-spectrum \cite{Kenton:2015lxa}. $f_{NL}$ is related to the amplitude of the bi-spectrum $A_\Rc\LF k_1,\,k_2,\,k_3 \RF$ as 
\begin{equation}
	f_{NL} = -\frac{5}{6}\frac{A_\Rc\LF k_1,\,k_2,\,k_3 \RF}{\sum_i k_i^3}\,, 
	\label{fnldef}
\end{equation}
where $A_\Rc\LF k_1,\,k_2,\,k_3 \RF$ stands for the redefinition of the bi-spectrum $B_\Rc$: 
\begin{equation}
	B_{\Rc}\left(k_{1},k_{2},k_{3}\right) = 4\pi^4\frac{1}{\prod_i k_i^3} \Pc_{\Rc}^{ 2} A_{\Rc}\LF k_1,\,k_2,\,k_3 \RF \, .
	\label{Bisp}
\end{equation}
{Combining \eqref{fnldef}, \eqref{Bisp} with \eqref{3corr} we obtain 
\begin{equation}
	\langle\Rc\left(\mathbf{k_{1}}\right)\Rc\left(\mathbf{k_{2}}\right)\Rc\left(\mathbf{k_{3}}\right)\rangle  =  \left(2\pi\right)^{7}\delta^{3}\left(\mathbf{k_{1}}+\mathbf{k_{2}}+\mathbf{k_{3}}\right) \frac{\sum_i k_i^3}{\prod_i k_i^3}  \Bigg(-\frac{3}{10} f_{NL}\LF k_1,\,k_2,\,k_3 \RF \Pc_{\Rc}^{ 2} \Bigg),
\end{equation}
which is the standard form of 3-point correlation widely used in the literature \cite{Chen:2006xjb,Chen:2006nt,Babich:2004gb,Seery:2005wm,komatsu2009nongaussianity} and also it is the definition that is used in the Planck data \cite{Akrami:2019izv}. We note that the same definition is used in our earlier work on PNG in the context of $R^2$-like inflation in quadratic curvature non-local theory \cite{Koshelev:2020foq}.}

Here $\Pc_R$ is the power spectrum of curvature perturbation \cite{Koshelev:2022olc}
\begin{equation}
	\Pc_{\Rc} \approx \frac{1}{3f_0\bar{R}_{\rm dS}} \frac{H^2}{16\pi^2\epsilon^2}
	\end{equation}
	where $H$ is the Hubble parameter during inflation and $\epsilon = -\frac{\dot{H}}{H^2}\approx \frac{1}{2N}$ with $N\gg 1$ where $N$ represents the number of e-folds. 
To calculate $f_{\rm NL}$, we consider the third order variation of \eqref{NAID} around the background \eqref{mR2in} which can be computed as  
\begin{equation}
	\delta_{(s)}^{(3)}S_H^{\rm Non-local} =   \delta_{(s)}^{(3)}S_{R+R^2}^{\rm local}+\delta_{(s)}^{(3)}S_{R+R^2}^{\rm Non-local} + \delta_{(s)}^{(3)}S_{\mathbb{R}^3}^{\rm Non-local}
	\label{3rdv}
\end{equation}
where 
\begin{equation}
	\begin{aligned}
		S_{R+R^2}^{\rm local}  = & \int d^4x\sqrt{-g}\Bigg[\frac{M_p^2}{2}R+ \frac{f_0}{2} R^2 \Bigg] \\ 
		S_{R+R^2}^{\rm Non-local} = &  \int d^4x\sqrt{-g}\Bigg\{\frac{M_p^2}{2}R+ \frac{1}{2} R\Bigg[\Fc_{R}\LF \square_s \RF -f_0\Bigg] R \Bigg\} \\ 
		S_{R^3}^{\rm Non-local} = &  \int d^4x\sqrt{-g}\Bigg[\Lc_1\LF \square_s\RF R\Lc_2\LF \square_s\RF R\Lc_3\LF \square_s \RF R \Bigg] \\
	\end{aligned}
	\label{listact}
\end{equation}
and the subscript $_{(s)}$ in \eqref{3rdv} denotes that we only consider 3-point scalar correlations, so we are dropping all interactions containing tensor modes.  
Using the calculations performed in \cite{Koshelev:2020foq}, since $\Phi+\Psi\approx 0 $ during inflation and the variation $\delta_{(s)}W_{\mu\nu\rho\sigma}\propto \Phi+\Psi$, we can conclude that all the terms involving Weyl tensor do not contribute to the scalar PNGs. In \eqref{NAID}, we can further consider quartic order non-local scalar curvature term:
\begin{equation}
	S^{\rm Non-local}_{R^4} = \frac{f_0\lambda_q}{2\Mc_s^4}\int d^4x\sqrt{-g}\Bigg[ \Lc_4 \LF \square_s \RF R \Lc_1 \LF \square_s \RF R \Lc_2 \LF \square_s \RF R \Lc_2 \LF \square_s \RF R\,  \Bigg]. 
	\label{r4nl}
\end{equation}
Here $\Lc_4\LF \square_s \RF$ is an arbitrary analytic infinite derivative operator. It is easy to deduce that \eqref{r4nl} still admits the inflationary solution \eqref{mR2in}. Applying  $\Lc_i\LF \frac{M^2}{\Mc_s^2} \RF =0$ ($i=1,2,3$), we can conclude that
the second order variation of \eqref{r4nl} around the background satisfying \eqref{mR2in} becomes zero exactly 
\begin{equation}
	\delta^{(2)}S_{R^4}^{\rm Non-local} =0\,, 
\end{equation}
whereas the 3rd order variation of \eqref{r4nl} around \eqref{mR2in} in the leading order de Sitter approximation is 
\begin{equation}
	\delta^{(3)}S_{R^4}^{\rm Non-local} \approx \Lc_4\LF \frac{M^2}{\Mc_s^2} \RF\frac{\lambda_q \bar{R}_{\rm dS}}{\lambda_c\Mc_s^2}  \delta^{(3)}S_{R^3}^{\rm Non-local} \,. 
\end{equation}
Given we assume 
\begin{equation}
	\Lc_4\LF \frac{M^2}{\Mc_s^2} \RF\frac{\lambda_q \bar{R}_{\rm dS}}{\lambda_c\Mc_s^2}  \ll 1\, ,
\end{equation}
we can neglect the contribution of \eqref{r4nl} to scalar (3-point) PNGs. However, we can possibly have non-negligible contributions to the trispectrum (or) 4-point correlations due to this term. The same logic can be easily extended to other higher scalar curvature non-local terms. Similarly, we can add quartic in Weyl tensor terms to \eqref{NAID} which do not contribute to the 3-point inflationary correlations due to the fact that background Weyl tensor vanishes in FLRW.

Therefore, non-local contributions to the bispectrum arise only from the part of the action \eqref{NAID} which is quadratic and cubic in Ricci scalar. PNGs from the non-local quadratic term in Ricci scalar were computed in \cite{Koshelev:2020foq}. We borrow and present here these results with two important changes. First, we include the contributions that explicitly break the scale invariance of the bispectrum. These contributions are important for studying the running of PNGs that we compute in the next section. Second, we present PNGs for the formfactors we introduced in \eqref{formfinal}. 
In Appendix~\ref{App3NG} we compute PNGs arising from the non-local cubic in Ricci scalar term which gives {several large} PNGs that we shall discuss in the next section. Finally, {the 3rd order action \eqref{3rdv} we obtain contains the following interactions of curvature perturbation $\Rc$} in the leading order slow-roll approximation as\footnote{We have corrected here minor typographical errors in the expression (4.12) of \cite{Koshelev:2020foq}}
\begin{equation}
	\begin{aligned}
		\delta_{(s)}^{(3)}S_{H}^{\rm Non-local} = \, & 4\frac{M_p^2}{H^2}\epsilon  \int   d\tau d^3x \Bigg\{B_1 \tau^{-2} \Rc\nabla\Rc\cdot\nabla\Rc+
		B_2 \tau^{-2} \Rc\Rc^{\prime 2}+  B_3 	\tau^{-3} \Rc\Rc\Rc^\prime\\+&
		B_4 \tau^{-1}\Rc^{\prime 3}+ B_5 
\tau^{-4}\Rc^3+ B_6\tau^{-1}\nabla\Rc\cdot\nabla\Rc\Rc^\prime + B_7 \Rc^\prime \nabla\Rc\cdot\nabla\Rc^\prime  \Bigg\}\,, 
	\end{aligned}
	\label{3rda}
\end{equation}
where $B_1$ to $B_7$ are dimensionless parameters  (approximated to be constant during inflation) {that give the final amplitude of the bispectrum \eqref{ampl} after the computation of 3-point correlation following \eqref{3-point-f}} 
\begin{equation}
	\begin{aligned}
		B_1 & = -2\epsilon-\frac{3\epsilon^2}{4} \\ 
		B_2 & = 2\epsilon+ \frac{3\epsilon^2}{4} +\frac{16}{3}\epsilon \Tc_{\rm NL} + \frac{8}{3} \epsilon ^3 \frac{ \bar{R}_{\rm dS}^2 }{ \Mc_S^4} \gamma_S^\dagger\LF \frac{\bar{R}_{\rm dS}}{4\Mc_s^2} \RF e^{ \gamma_S\LF \frac{\bar{R}_{\rm dS}}{4\Mc_s^2} \RF  } -\frac{2\lambda_c}{9}\frac{\bar{R}_{\rm dS}}{\Mc_s^2}\epsilon\LF 2\epsilon^2T_{\rm NL}^3+\epsilon T^2_{\rm NL}+ T_{\rm NL}^1 \RF\\
		B_3 & = -32\Tc_{\rm NL}-\frac{8\lambda_c}{9}\frac{\bar{R}_{\rm dS}}{\Mc_s^2}\epsilon\LF 2\epsilon T_{\rm NL}^2+T_{\rm NL}^1 \RF \\ 
		B_4 & = -2 \Tc_{\rm NL} -\frac{1}{54}\frac{\bar{R}_{\rm dS}}{\Mc_s^2}\epsilon\LF 8\epsilon^3 T_{\rm NL}^4+4\epsilon^2T_{\rm NL}^3+2\epsilon T_{\rm NL}^2+T_{\rm NL}^1 \RF \\ 
		B_5 & = -\frac{\epsilon^2}{2}+\frac{32\bar{R}_{\rm dS}}{\Mc_s^2}\epsilon T_{\rm NL}^1\,\\
		B_6 & = -2\Tc_{\rm NL}\\
		B_7 & =  \frac{16}{3}\epsilon \Tc_{\textrm{NL}}+\frac{8}{3} \epsilon ^3 \frac{ \bar{R}_{\rm dS}^2 }{ \Mc_S^4} \gamma_S^\dagger\LF \frac{\bar{R}_{\rm dS}}{4\Mc_s^2} \RF e^{ \gamma_S\LF \frac{\bar{R}_{\rm dS}}{4\Mc_s^2} \RF  } \, , 
	\end{aligned} 
\end{equation}
where  the subscript \say{$_{\rm dS}$} denotes the quantities in quasi-dS approximation and
\begin{equation}
	\Tc_{\rm NL} = \frac{1}{3}\epsilon ^2\left(e^{\gamma_S\LF \frac{\bar{R}_{\rm dS}}{4\Mc_s^2} \RF }-1\right) +\epsilon ^3 \frac{ \bar{R}_{\rm dS}^2 }{12 \Mc_S^4} \gamma_S^\dagger\LF \frac{\bar{R}_{\rm dS}}{4\Mc_s^2} \RF e^{ \gamma_S\LF \frac{\bar{R}_{\rm dS}}{4\Mc_s^2} \RF  } 
\end{equation}
 And the quantities $T_{\rm NL}^1,\cdots, T_{\rm NL}^4$ are defined in \eqref{tnls}.

Computing the amplitude of 3-point correlation, we obtain 
\begin{equation}
	A_\Rc = \sum_{i=1}^{7} B_iS_i \,, 
	\label{ampl}
\end{equation}
where 
\begin{equation}
	\begin{aligned}
		S_1 & =  2\boldsymbol{k}_1\cdot \boldsymbol{k}_2 \LT K-\frac{k_1k_2+k_2k_3+k_3k_1}{K} -\frac{k_1k_2k_3}{K^2} \RT +\textrm{perms}\, ,\\
		S_2 &  = \frac{2k_1^2k_2^2}{K} +\frac{2k_1^2k_2^2k_3}{K^2}+\textrm{perms}\,, \\
		S_3 &  \approx \,   k_3^2\LT -2K-\frac{2k_1k_2}{K}  \RT+\Cc\LF z \RF k_3^3 +\text{perms}\,\\
	\end{aligned}
\end{equation}
\begin{equation}
	\begin{aligned}
		S_4 &  = \frac{4k_1^2k_2^2k_3^2}{K^3}+\text{perms} \\
		S_5 &  =  - \frac{K^3}{3} +2Kk_1k_2+\frac{k_1k_2k_3}{3}  +\text{perms}\\
		S_6 & =  2\LF \boldsymbol{k}_1\cdot \boldsymbol{k}_2 \RF k_3^2\LT \frac{2}{K}+\frac{2k_1+2k_2}{K^2}+\frac{4k_1k_2}{K^3} \RT +\text{perms}\\
		S_7& =   \LF \boldsymbol{k}_2\cdot \boldsymbol{k}_3 \RF k_1^2k_3^2\LT -\frac{2}{K^3}-\frac{6k_2}{K^4} \RT 
	\end{aligned}
\end{equation}
where $z = \frac{K}{K_\ast}$ with $K_\ast =a_\ast H_\ast = 0.05\, {\rm Mpc}^{-1}$ is a particular reference scale and $C(z) \approx \gamma_E+\ln z -\frac{z^2}{4}+\frac{z^4}{96}$. 
In deriving \eqref{3rda}, we used the following on shell relations which are the result of the background solution \eqref{mR2in} and the equation of motion for curvature perturbation at the linearized level \cite{Koshelev:2020foq,SravanKumar:2019eqt}:
\begin{equation}
	\begin{aligned}
		\bar\square_s\Rc\approx &\, \frac{M^2}{\Mc_{s}^2}\Rc \implies & \Oc\LF\bar\square_s\RF \Rc \approx &\,\Oc\LF \frac{M^2}{\Mc_{s}^2} \RF\Rc \,\\
		\bar\square_s\Rc^{\prime}\approx &\, \LF\frac{\bar\square_s}{\Mc_s^2}+\frac{\bar{R}_{\rm dS}}{4\Mc_s^2}\RF \Rc^\prime \implies & \Oc\LF\bar\square_s\RF \Rc^\prime \approx &\, \Oc\LF \frac{M^2}{\Mc_{s}^2}+\frac{\bar{R}_{\rm dS}}{4\Mc_{s}^2} \RF\Rc^\prime\,, 
	\end{aligned}
	\label{Commtr}
\end{equation}
where $\Oc$ is an arbitrary analytic operator.  The above on-shell relations together with \eqref{mR2in} bring the 3rd order action of the non-local gravity \eqref{NAID} into the local form, and all the effect of non-localities is transferred into the on-shell vertex factors $\Tc_{\rm NL}$ and $T_{\rm NL}^1,\cdots,T_{\rm NL}^4$. 

{From \eqref{3rda} we can notice that the first two interaction terms of curvature perturbation are the standard leading order interactions of the local $R^2$ model and also of the canonical single field inflation. In the case of non-local $R^2$-like inflation, the second term in  \eqref{3rda} gets non-local contributions. 
	The 3rd term in \eqref{3rda} is actually the higher order slow-roll term that appears in the standard canonical single field inflation (See (3.12) of \cite{Chen:2006xjb}) but here in our case the contribution of this term can become significant due to the formfactors present in our generalized non-local gravity \eqref{NAID}. Going to the second line of \eqref{3rda} we can again identify the first three terms which are very much analogous to  the interactions of curvature perturbation in the standard canonical case \cite{Chen:2006xjb} but here in the expressions from $B_3$ to $B_6$ we have the nonlocal factors which can win over the slow-roll parameters and ultimately leading to large PNGs. Also, we can notice that interaction terms of these kind especially the term proportional to $B_4$ is one of the dominant contribution in the context of non-canonical single field inflation for small sound speed ($c_s$) \cite{Chen:2006nt}. But here in our case the effect of this interaction term is due to nonlocality rather than sound speed which happens to be Unity in our case.
	The last term in the second line of \eqref{3rda} appears to be a new interaction in comparison with local theories of general single field inflation \cite{Chen:2006xjb,Chen:2006nt}. However, in our analysis, we find the contribution of this term to be negligible since this term is found to be sub-dominant compared to the remaining interaction terms with scale dependent vertex factors $B_i$. Note that in the limit $\Mc_s\to \infty$ (local limit) we recover the known result for PNGs in standard canonical single field inflation.  }

\subsection{{Squeezed, equilateral and orthogonal limits} of $f_{\rm NL}\LF k_1,\,k_2,\,k_3 \RF$ and their running $n_{\rm NG}$} 

\label{sec:scaleNG}

In this section, we explore {the most observationally targeted limits of} the reduced bispectrum  $f_{\rm NL}\LF k_1,\,k_2,\,k_3 \RF$ and their running \cite{Kenton:2015lxa,Chen:2006nt,Fergusson:2008ra} which are very significant observables to probe the nature of the scalar field (either scalaron or inflaton) and its interactions. Discussion of results obtained in this section are presented in Sec.~\ref{sec:results} and Sec.~\ref{sec:eftcom}. 
The PNGs can quantified by plotting $f_{\rm NL}\LF k_1,\,k_2,\,k_3 \RF$ in terms of the following redefinition of wave-numbers \cite{Babich:2004gb,Fergusson:2008ra}:
\begin{equation}
	k_1 = \frac{K}{4} \left(\alpha _s+\beta _s+1\right),\quad k_2 =\frac{K}{4} \left(-\alpha _s+\beta _s+1\right),\quad k_3 = \frac{K}{2} \left(1-\beta _s\right)\, .
	\label{alsbs}
\end{equation}
If future observations will probe fully the functional dependence of $f_{\rm NL}\LF K,\,\alpha_s,\,\beta_s \RF$, we can fully determine self-interactions of the scalarons we see in \eqref{3rdv} which are by nature not scale-invariant.  PNGs of the generalized non-local $R^2$-like inflation \eqref{NAID} are computed by substituting \eqref{ampl} in \eqref{fnldef}. 
There are three {limits} of reduced bi-spectrum called \say{equilateral: $k_1=k_2=k_3=k/3$}, \say{orthogonal: $4k_1=4k_2= 2k_3=k$}, \say{squeezed: $k_3\ll k_1=k_2=\frac{k}{2}$} (where the total momentum $K=k$) which are more relevant popular templates with the present CMB  constraints \cite{Akrami:2019izv}.
Computing $f_{\rm NL}$ for these three configurations, we obtain 
\begin{equation}
	\begin{aligned}
		f_{\rm NL }^{\rm sq}  \approx & \, \frac{5}{12} \LF 1-n_s\RF -35.5\,\Tc_{\rm NL}+8.9\,\Cc(z)\,\Tc_{\rm NL} -1.1\,\epsilon^3\,\frac{\bar{R}^2_{\rm dS}}{\Mc_s^4}\,\gamma_S^\dagger\LF \frac{\bar{R}_{\rm dS}}{4\Mc_s^2} \RF \\  &-\lambda_c\,\frac{\bar{R}_{\rm dS}}{\Mc_s^2}\, \Bigg( 5.8 \,\epsilon^2\,T_{\rm NL}^2 -  1.5\, \Cc\LF z \RF   \epsilon^2\, T_{\rm NL}^2 - 0.19 \,\epsilon^3\,  T_{\rm NL}^3\Bigg) \,,\\
		f_{\rm NL }^{\rm equiv}  \approx & \, \frac{5}{12} \LF 1-n_s\RF -46.6\,\Tc_{\rm NL}+8.9\,\Cc(z)\,\Tc_{\rm NL} -1.8\,\epsilon^3 \,\frac{\bar{R}^2_{\rm dS}}{\Mc_s^4}\,\gamma_S^\dagger\LF \frac{\bar{R}_{\rm dS}}{4\Mc_s^2} \RF \\& -\lambda_c\frac{\bar{R}_{\rm dS}}{\Mc_s^2}\, \Bigg(7.7 \,\epsilon^2 \, T_{\rm NL}^2 - 1.5\, \Cc\LF z \RF  \, \epsilon^2\, T_{\rm NL}^2- 0.3\,\epsilon^3 \, T_{\rm NL}^3-0.02\,\epsilon^4\,  T_{\rm NL}^4\Bigg) \, ,\\
		f_{\rm NL }^{\rm ortho}  \approx & \, \frac{5}{12} \LF 1-n_s\RF -39.1\,\Tc_{\rm NL}+8.9\,\Cc(z)\,\Tc_{\rm NL} -1.2\,\epsilon^3\,\frac{\bar{R}^2_{\rm dS}}{\Mc_s^4}\,\gamma_S^\dagger\LF \frac{\bar{R}_{\rm dS}}{4\Mc_s^2} \RF \\& -\lambda_c\frac{\bar{ R}_{\rm dS}}{\Mc_{s}^2}\, \Bigg( 6.4\, \epsilon^2 \, T_{\rm NL}^2- 1.5\,  \Cc\LF z \RF \,  \epsilon^2 T_{\rm NL}^2 -0.2\, \epsilon^3\,  T_{\rm NL}^3-0.01\, \epsilon^4\, T_{\rm NL}^4 \Bigg) \, .
	\end{aligned}
	\label{fnlseo}
\end{equation}
From \eqref{fnlseo} we can conclude that all the crucial $f_{\rm NL}$ configurations get non-local contributions. 
There are three types of non-local contributions appearing in the expressions \eqref{fnlseo}. 
\begin{enumerate}
	\item Contributions involving $e^{\gamma_S\LF \square_s \RF}$ come from the term that is quadratic in scalar curvature \eqref{listact} for the formfactor $\Fc_R\LF \square_s \RF$ \eqref{formfinal}.
	\item Contribution from the cubic non-local scalar curvature term \eqref{listact}:  this can be read from the terms involving $T_{\rm NL}^2-T_{\rm NL}^4$. As explained in Appendix~\ref{App3NG}, we found the contributions of the terms involving $T_{\rm NL}^1$ are negligible compared to other terms and therefore we drop them in \eqref{fnlseo} for brevity.
	\item Contribution involving $\Cc\LF \frac{K}{K_\ast}\RF$ which comes from taking carefully the infrared (IR) limit of integration using the judicious choice $\tau= - \frac{1}{K_\ast}$ \cite{Seery:2010kh,Pajer:2016ieg}.  Usually this contribution is slow-roll suppressed in standard single field models of inflation \cite{Pajer:2016ieg,Chen:2006nt,Burrage:2011hd}, but in our case it is modulated by analytic non-local contributions\footnote{In our result \eqref{fnlseo} we neglected small local contributions of the order $O\LF \epsilon^2  \RF$. } when $\bar{R}\gtrsim \Mc_s^2$. 
\end{enumerate}
From \eqref{fnlseo} we can deduce that the non-Gaussianity parameter $f_{\rm NL}$ that can be potentially observable in future depends on the following set of quantities
\begin{equation}
	\Bigg\{e^{\gamma_S\LF \frac{\bar{R}_{\rm dS}}{4\Mc_s^2}\RF},\,\gamma^{\dagger}_S\LF \frac{\bar{R}_{\rm dS}}{4\Mc_s^2}\RF,\,\lambda_c,\, e^{\ell_i\LF \frac{\bar{R}_{\rm dS}}{4\Mc_s^2} \RF} \Bigg\}\, .
	\label{parameterNG}
\end{equation}
As we discussed in the previous section, the scale invariance of bispectrum \eqref{ampl} is broken in the non-local $R^2$-like inflation due to the scale dependence of quantities \eqref{parameterNG} through 
\begin{equation}
	\Bigg\{\bar{R}_{\rm dS}\LF k \RF,\quad \Cc\LF k \RF\Bigg\}
	\label{sdrc}
\end{equation}
Due to the presence of exponentials, we can expect that the scale dependence can be significant. To quantify it we define a new parameter called running of $f_{\rm NL}$ as \cite{Chen:2006nt} 
\begin{equation}
	n_{\rm NG} \equiv \frac{d\ln f_{\rm NL}}{d\ln k}
	\label{nfnl}
\end{equation}
which we can evaluate in the various limits such as squeezed, equilateral and orthogonal. The $n_{\rm NG}$ can be evaluated using the following quantities 
\begin{equation}
	\begin{aligned}
		\frac{d f_{\rm NL }^{\rm sq} }{d\ln k} \approx & \, -\frac{dn_s}{d\ln k} +\Bigg(  -35.5+8.9\,\Cc(z) \Bigg) \frac{\pd \Tc_{\rm NL}}{\pd N} + \Cc^\dagger \LF z \RF\Bigg( 8.9\,  \Tc_{\rm NL} +\epsilon^2\lambda_c\frac{\bar{R}_{\rm dS}}{\Mc_s^2} T_{\rm NL}^2 \Bigg)\\ & -2.2\,\epsilon^4\,\frac{\bar{R}^3_{\rm dS}}{\Mc_s^6}\Bigg[\gamma_S^\dagger\LF \frac{\bar{R}_{\rm dS}}{4\Mc_s^2} \RF +\frac{1}{4} \gamma_S^{\dagger 2}\LF \frac{\bar{R}_{\rm dS}}{4\Mc_s^2} \RF +\frac{1}{4} \gamma_S^{\dagger \dagger}\LF \frac{\bar{R}_{\rm dS}}{4\Mc_s^2} \RF\Bigg] \\  &-\epsilon\,\lambda_c \frac{2\bar{R}_{\rm dS}}{\Mc_s^2} \Bigg( 5.8 \,\epsilon^2\, T_{\rm NL}^2 -  1.5\,  \epsilon^2\,\Cc\LF z \RF   T_{\rm NL}^2 - 0.76\,\epsilon^3  T_{\rm NL}^3\Bigg) \\ &
		+\epsilon\,\lambda_c \frac{\bar{R}^2_{\rm dS}}{2\Mc_s^4} \Bigg( 5.8 \,\epsilon^2\,\frac{\pd T_{\rm NL}^2}{\pd N} -  1.5\, \epsilon^2\,\Cc\LF z \RF \,   \frac{\pd T_{\rm NL}^2}{\pd N}- 0.19\,\epsilon^3 \, \frac{\pd T_{\rm NL}^3}{\pd N}\Bigg) \, ,\\
	\end{aligned}
\end{equation}
\begin{equation}
	\begin{aligned}
		\frac{df_{\rm NL }^{\rm eq}}{d\ln k}  \approx & \, -\frac{dn_s}{d\ln k} +\Bigg(  -46.6+8.9\,\Cc(z) \Bigg) \frac{\pd \Tc_{\rm NL}}{\pd N}+  \Cc^\dagger \LF z \RF\Bigg( 8.9\,  \Tc_{\rm NL} +\epsilon^2\,\lambda_c\,\frac{\bar{R}_{\rm dS}}{\Mc_s^2}\, T_{\rm NL}^2 \Bigg) \\&
		-3.6\,\epsilon^4\,\frac{\bar{R}^3_{\rm dS}}{\Mc_s^6}\Bigg[\gamma_S^\dagger\LF \frac{\bar{R}_{\rm dS}}{4\Mc_s^2} \RF +\frac{1}{4} \gamma_S^{\dagger 2}\LF \frac{\bar{R}_{\rm dS}}{4\Mc_s^2} \RF +\frac{1}{4} \gamma_S^{\dagger \dagger}\LF \frac{\bar{R}_{\rm dS}}{4\Mc_s^2} \RF\Bigg]\\&
		-\epsilon\,\lambda_c\frac{2\bar{R}_{\rm dS}}{\Mc_s^2} \Bigg(7.7 \,\epsilon^2 \, T_{\rm NL}^2 - 1.5\,\epsilon^2\, \Cc\LF z \RF    T_{\rm NL}^2- 1.2\, \epsilon^3\,  T_{\rm NL}^3-0.12\,\epsilon^4\, T_{\rm NL}^4\Bigg) \\ &
		+\epsilon\,\lambda_c\frac{\bar{R}^2_{\rm dS}}{2\Mc_s^4} \Bigg(7.7 \epsilon^2 \frac{\pd T_{\rm NL}^2}{\pd T_{\rm NL}^2} - 1.5\,\epsilon^2\, \Cc\LF z \RF   \, \frac{\pd T_{\rm NL}^2}{\pd N}- 0.3\,\epsilon^3 \, \frac{\pd T_{\rm NL}^3}{\pd N}-0.02\,\epsilon^4  \,\frac{\pd T_{\rm NL}^4}{\pd N}\Bigg) \, , \\
		%running orth
		\frac{df_{\rm NL }^{\rm orth}}{d\ln k}  \approx & \, -\frac{dn_s}{d\ln k} +\Bigg(  -39.1+8.9\,\Cc(z) \Bigg) \frac{\pd \Tc_{\rm NL}}{\pd N}+  \Cc^\dagger \LF z \RF\Bigg( 8.9\,  \Tc_{\rm NL} +\epsilon^2\lambda_c\frac{\bar{R}_{\rm dS}}{\Mc_s^2} T_{\rm NL}^2 \Bigg) \\&
		-2.4\,\epsilon^4\,\frac{\bar{R}^3_{\rm dS}}{\Mc_s^6}\Bigg[\gamma_S^\dagger\LF \frac{\bar{R}_{\rm dS}}{4\Mc_s^2} \RF +\frac{1}{4} \gamma_S^{\dagger 2}\LF \frac{\bar{R}_{\rm dS}}{4\Mc_s^2} \RF +\frac{1}{4} \gamma_S^{\dagger \dagger}\LF \frac{\bar{R}_{\rm dS}}{4\Mc_s^2} \RF\Bigg]\\&
		-\epsilon\,\lambda_c\frac{2\bar{R}_{\rm dS}}{\Mc_s^2} \Bigg(6.4 \,\epsilon^2\, T_{\rm NL}^2 - 1.5\,\epsilon^2 \Cc\LF z \RF  \,  T_{\rm NL}^2- 0.8\,\epsilon^3\,  T_{\rm NL}^3-0.06\,\epsilon^4  \, T_{\rm NL}^4\Bigg) \\ &
		+\epsilon\,\lambda_c\frac{\bar{R}^2_{\rm dS}}{2\Mc_s^4} \Bigg(6.4\, \epsilon^2 \, \frac{\pd T_{\rm NL}^2}{\pd N} - 1.5\, \Cc\LF z \RF  \, \epsilon^2 \, \frac{\pd T_{\rm NL}^2}{\pd N}- 0.2\,\epsilon^3\,  \frac{\pd T_{\rm NL}^3}{\pd N}-0.01\,\epsilon^4 \, \frac{\pd T_{\rm NL}^4}{\pd N}\Bigg)\, .
	\end{aligned}
	\label{Rfnlseo}
\end{equation}
From \eqref{Rfnlseo} we can deduce that running of PNGs depend on the higher derivatives of our formfactors \eqref{parameterNG}. This reveals that if we measure running of PNGs in future observations, we can probe the formfactors, reconstruct our theory and ultimately learn about UV-completion of gravity. 

\subsection{Quantum fluctuations and the bi-spectrum} 

In \cite{Koshelev:2022olc} it was shown that the sound speed of curvature perturbations is Unity in the non-local gravity inflation \eqref{NAID}.  Therefore, we have an effective scalar field (scalaron) propagating during inflation similar to the local $R^2$ theory. Moreover, it was shown in \cite{Koshelev:2022olc} that there is only a single scalar mode that is propagating during inflation. Thus, we can also conclude that the leading mode of the curvature perturbations $\Rc$, corresponding to the growing mode of energy density perturbations, is approximately time-independent on super-Hubble scales while the generic solution for $\Rc$ there has the form \cite{DeFelice:2010aj}
\begin{equation}
	\Rc = c_1 +c_2\int \frac{dt}{a^3\epsilon}\,. 
\end{equation}
This implies all the PNGs which are generated in the generalized non-local $R^2$-like inflation are determined by interactions of quantum fluctuations which evolve until the moment of the Hubble radius crossing during inflation that is exactly what we compute here. Obviously since all the modes become frozen on super-Hubble scales $k\ll aH$, the $f_{\rm NL}$ are solely determined by quantum field interactions when $k\sim aH$. 
We will see in the next sub-section that non-local self interactions of $\Rc$  can generate sizeable contributions for all $f_{NL}$ in squeezed, equilateral and orthogonal configurations when the non-locality scale is close to the Hubble scale during inflation $ \Mc_s\gtrsim H $. As we noted before, non-commutativity of covariant and partial  derivatives in the quasi-dS regime \eqref{Commtr} plays a crucial role in getting large PNGs in the generalized non-local $R^2$-like inflation. 
This result evades all known ways of obtaining detectable level of PNGs through non-canonical scalar field(s) and not adiabatic vacuum (or, non-Bunch-Davis) initial conditions \cite{Chen:2010xka}. Most notably, we modify the Maldacena consistency relation that was argued to take place for any general single field standard slow-roll inflation \cite{Creminelli:2004yq,Cheung:2007st,Burrage:2011hd}. We further discuss this in detail later in Sec.~\ref{sec:eftcom} and Appendix~\ref{app:violmc}.

\section{Numerical illustration of $f_{\rm NL}$ in various limits  and their running $n_{f_{\rm NL}}$}
\label{sec:NGRUN}
To illustrate our result with explicit numbers, we plot $f_{\rm NL}^{sq},\, f_{\rm NL}^{eq},\,f_{\rm NL}^{\rm orth}$ choosing the following entire functions $\ell_i\LF \square_{s} \RF$ that are compatible with \eqref{entic} 
\begin{equation}
	\gamma_S\LF \square_s \RF = \alpha_{1} \square_{s}\LF \square_{s}-\frac{M^2}{\Mc_s^2} \RF +\cdots,\quad 	\ell_i\LF \square_{s} \RF = \alpha_{i1} \square_{s}\LF \square_{s}-\frac{M^2}{\Mc_s^2} \RF+\cdots\,, 
	\label{chenc}
\end{equation}
where $\alpha_1,\,\alpha_{i1}$ are 4 dimensionless parameters. The $\cdots$ represent higher order terms which we assume to be irrelevant at the inflationary scales, but they are important to suppress formfactors in the infinite momentum limit $p\to\pm\infty$.  
The results we report here are not that much sensitive to the choice \eqref{chenc}, and in principle we can consider higher degree polynomials in $\square_s$ for $\gamma_S\LF \square_s \RF,\,\ell_i\LF \square_{s} \RF$, but choosing \eqref{chenc} we explicitly assume the effect of higher order terms in $\square_s$ to be negligible by setting the dimensionless coefficients of them $\ll \LF \alpha_1,\,\alpha_{i1} \RF$. 
We assume $\alpha_{i1}\sim O(1-10)$, so that we do not fine tune much our free parameters and we can rely on the scale of non-locality $\Mc_s$ for physics discussion. 

\subsection{The case with only non-local quadratic term in curvature scalar: $\lambda_c=0$}
\begin{figure}[h!]
	\centering
	\includegraphics[width=2.8in]{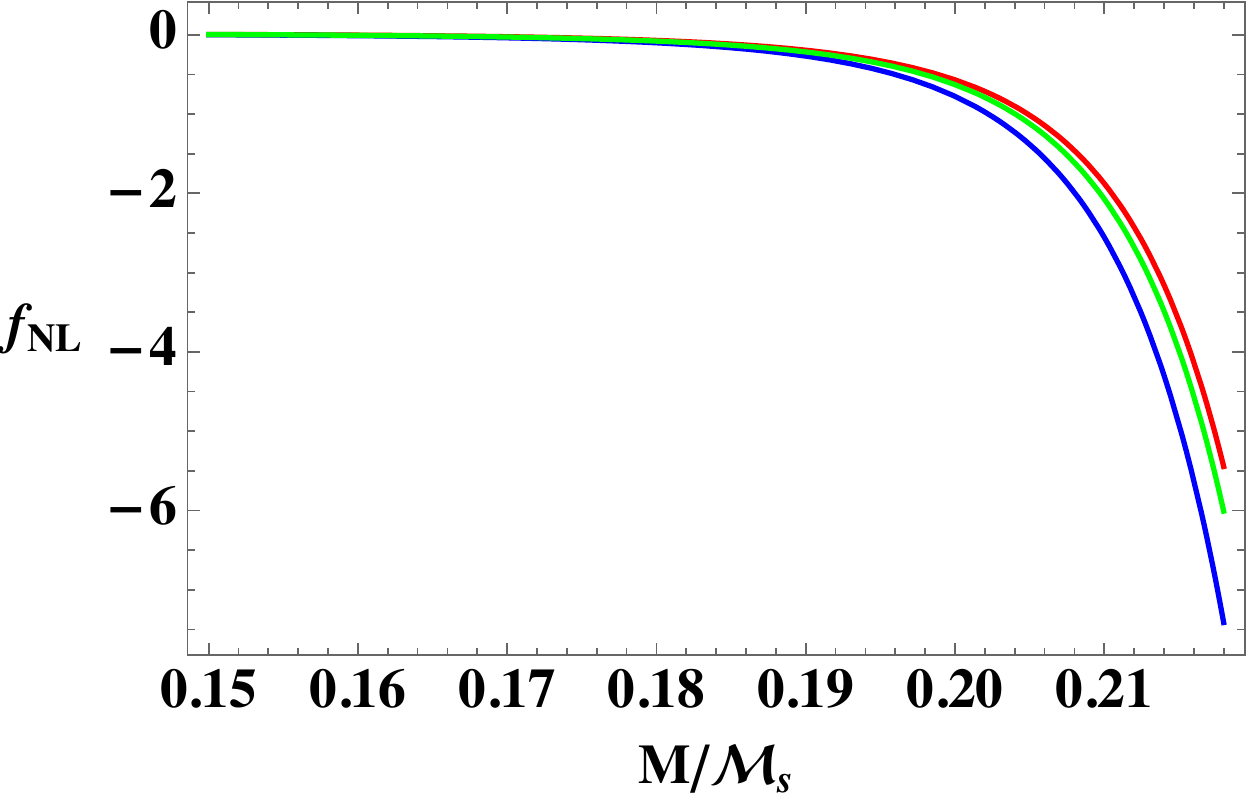}\quad \includegraphics[width=2.8in]{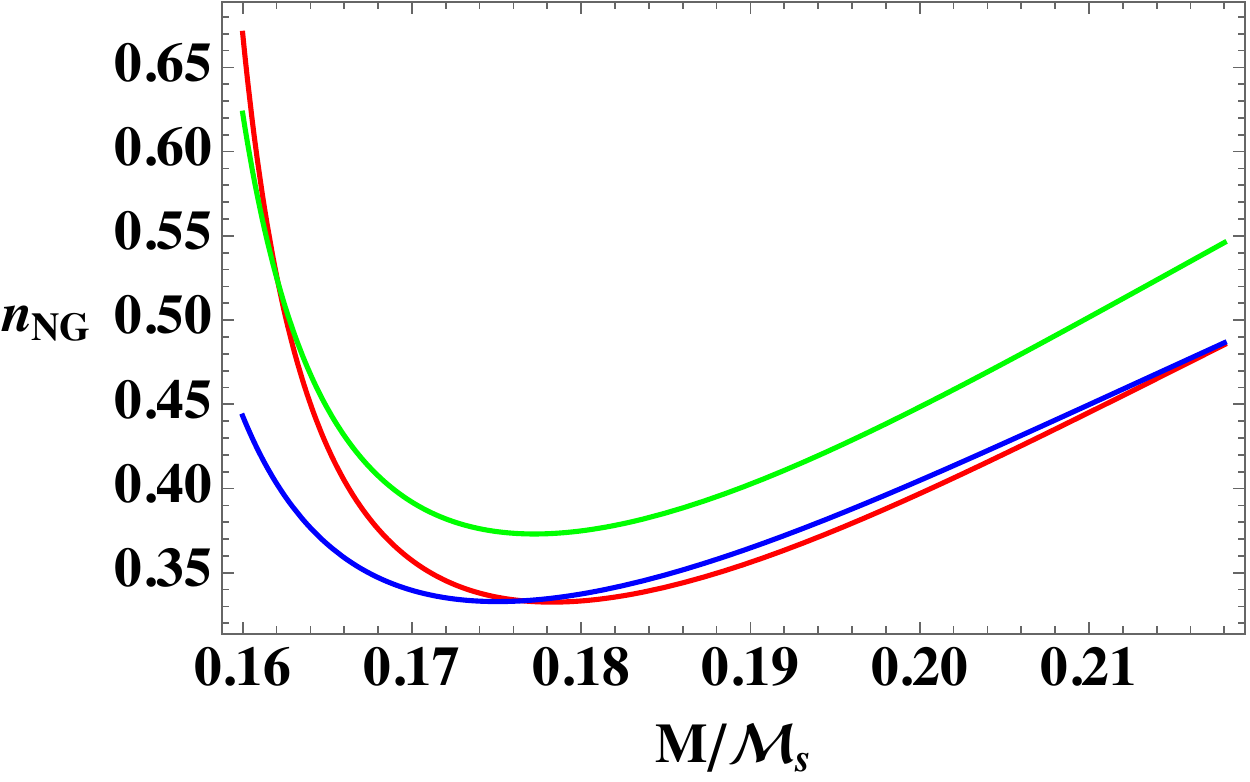}
	\caption{In the left panel we plot $f_{\rm NL}^{\rm sq},\,f_{\rm NL}^{\rm eq},\,f_{\rm NL}^{\rm orth}$ versus $M/\Mc_s$ (in red, blue and green coloured lines respectively) and in the right panel we plot the corresponding running of $f_{\rm NL}$. In both of the plots, we take $\alpha_1=1,\,\lambda_c=0$ and $N=55$ corresponding to the pivot scale $k_\ast=a_\ast H_\ast$. In these plots we recover the predictions of local $R^2$ gravity in the limit $\frac{M}{\Mc_s}\to 0$.}
	\label{fig:fnlqcg}
\end{figure}

In this section, we analyse the PNGs of a case without the cubic non-local Ricci scalar term in \eqref{NAID} by setting $\lambda_c =0$. Our results for the various {limits} of $f_{\rm NL}$ in this case lead to universal relations between them that do not explicitly depend on any choice of $\gamma_S\LF \square_s \RF$:
\begin{equation}
	f^{\rm eq}_{\rm NL}\approx 1.5f^{\rm sq}_{\rm NL},\quad f^{\rm orth}_{\rm NL}\approx 1.1f^{\rm sq}_{\rm NL}\,.
	\label{relQG}
\end{equation}
They can be verified through the expressions in \eqref{fnlseo} evaluating them at $k = k_\ast$, and when the non-local contribution dominates over $\frac{5}{12}\LF 1-n_s \RF$. We can also verify this numerically in Fig.~\ref{fig:fnlqcg} where we also plot the running of $f_{\rm NL}$ defined in \eqref{nfnl}. It is seen from it that \eqref{relQG} is satisfied while running of $f_{\rm NL}$ does not follow this relation because running depends on the quantity $\Cc^\dagger\LF z \RF$ explicitly. 

\subsection{PNGs with all the terms in \eqref{NAID}}
In this section, {we illustrate numerically the PNGs 
 in the generalized non-local $R^2$-like inflation by computing the values of $f_{\rm NL}$ in the squeezed, equilateral and orthogonal limits}. 
With $\lambda_c\neq 0$ we obtain new contributions for PNGs mimicking several classes of scalar field models  of inflation and EFTs discussed in Sec.~\ref{sec:results} and Sec.~\ref{sec:eftcom}. All of these {distinct possibilities of large $f_{\rm NL}$ in various limits are obtained by considering different values of parameters $\LF \alpha_1,\,\alpha_{i1} \RF$.}

\begin{enumerate}
	\item Large PNGs compatible with the Planck constraints   \eqref{boundsfnl}: see Fig.~\ref{R2CFNL}. 
	\begin{figure}[h!]
		\centering
		\includegraphics[width=2.8in]{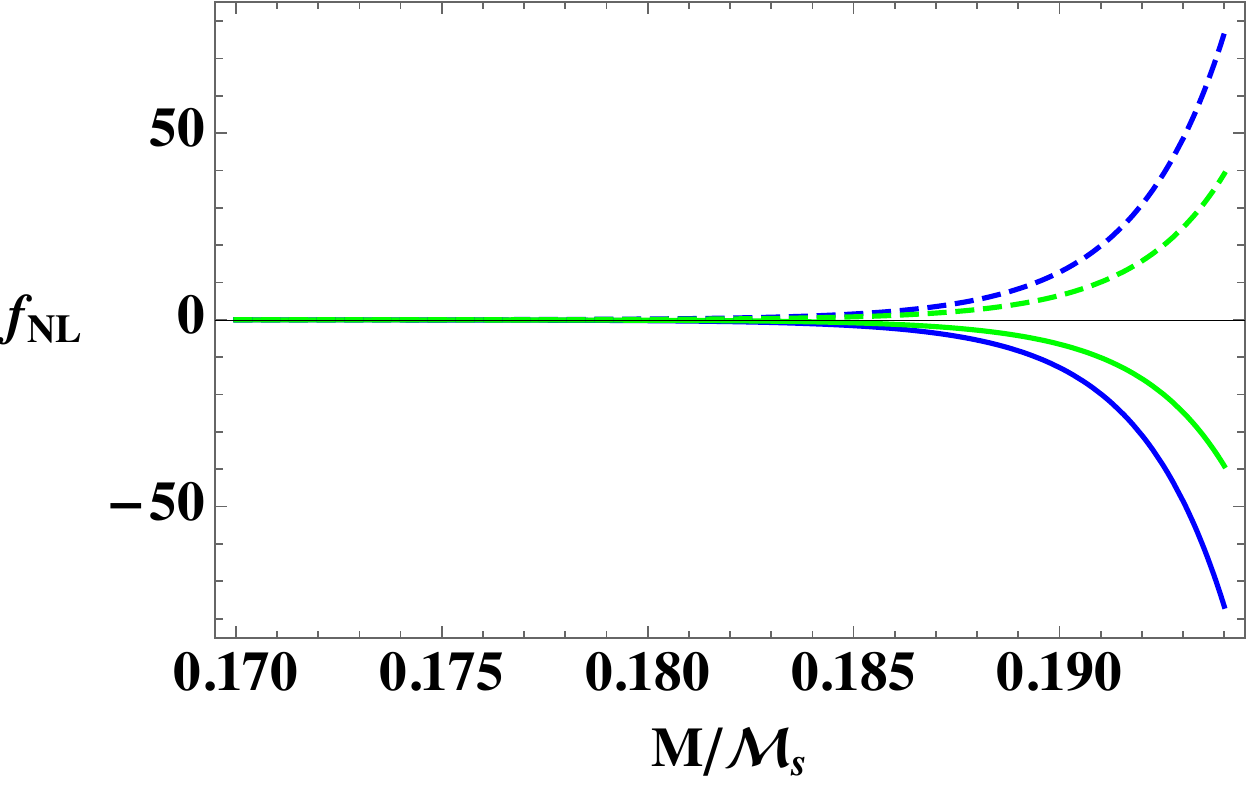}\quad \includegraphics[width=2.8in]{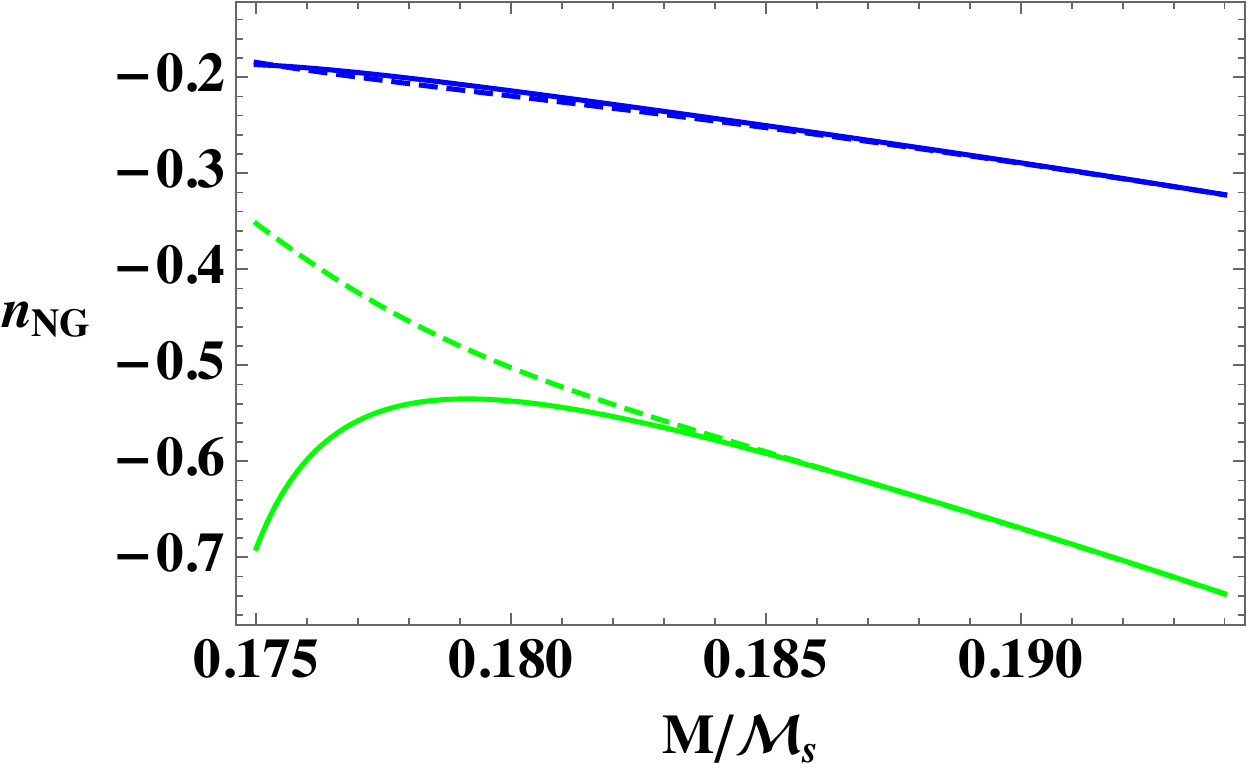}\quad  \includegraphics[width=2.8in]{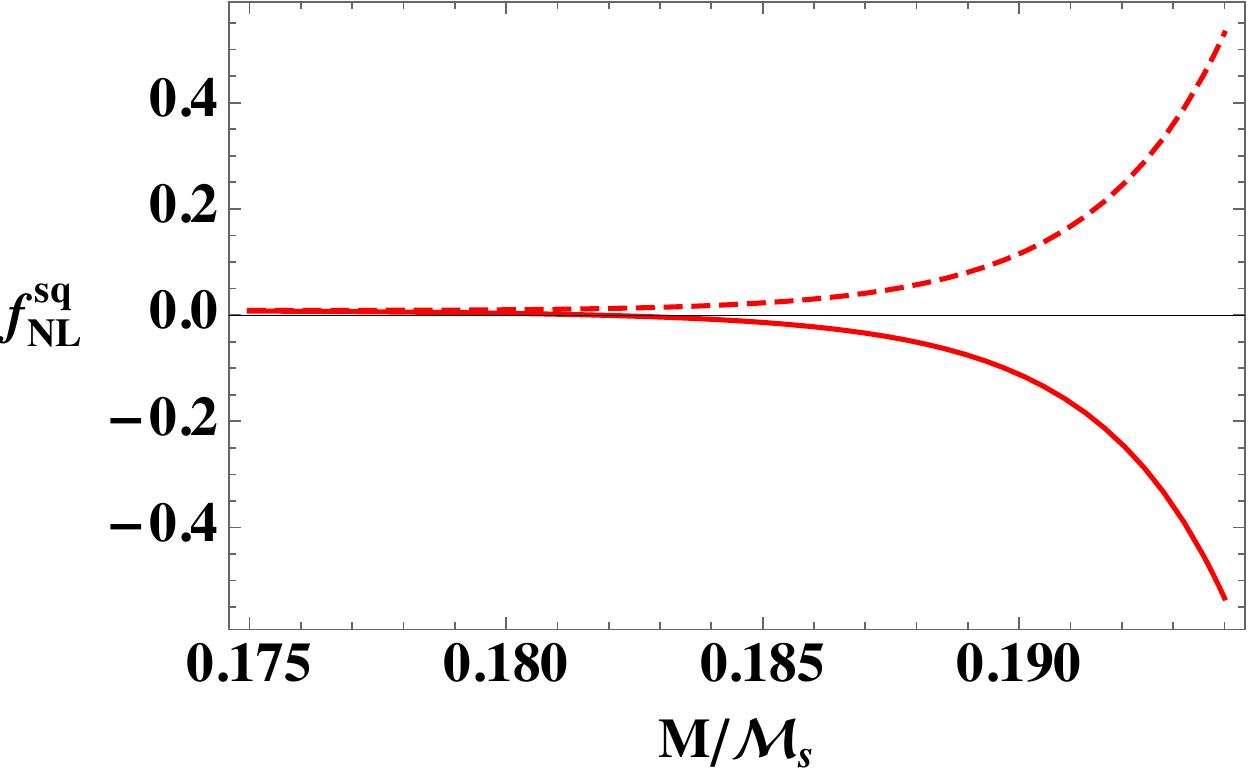}\quad \includegraphics[width=2.8in]{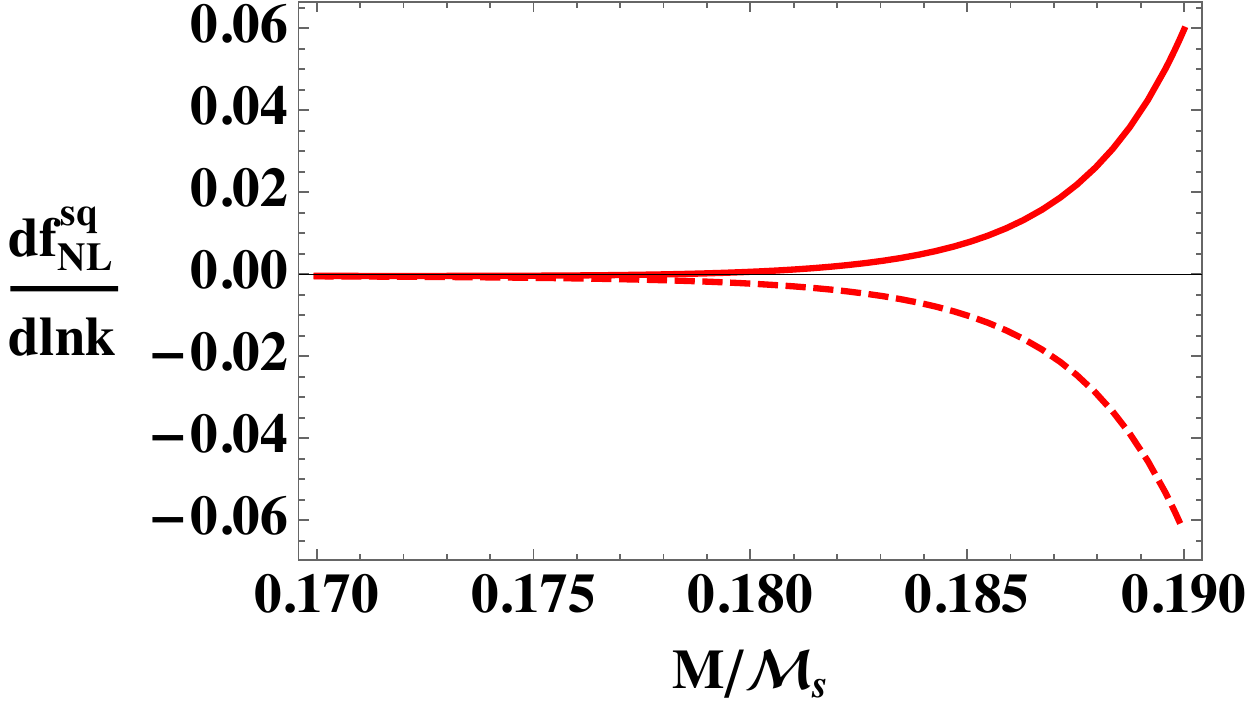}
		\caption{In all of the plots we present various {limits} of $f_{\rm NL}$ and their running $n_{\rm NG}$ versus $M/\Mc_s$ for  $\lambda_c>0$ (dashed lines) and $\lambda_c<0$ (full lines) respectively, while red, blue and green lines represent squeezed, equilateral and orthogonal {limits} respectively. For all of these plots we fix the parameters as $\lambda = \pm 4,\,\alpha_1=0.4,\,\alpha_{11}=3.1,\,\alpha_{21}= \alpha_{31}=1$ and $N=55$ corresponding to the pivot scale $k_\ast=a_\ast H_\ast$. }
		\label{R2CFNL}
	\end{figure}
	\item Large PNGs in the squeezed and orthogonal limit while negligible in the equilateral limit: see Fig.~\ref{R2zsor}. 
	
	\item Large PNGs in the equilateral and orthogonal limit while negligible in the squeezed limit: see Fig.~\ref{R2zsq}. {The predictions of Fig.~\ref{R2zsq} are very much similar to the signature of PNGs in the non-canonical single field inflation. Thus, one can expect the shape of $f_{NL}\LF k_1,\,k_2,\,k_3 \RF$ would be close to equilateral shape. But we must not forget the strong running of $f_{NL}^{eq}$ depicted in Fig.~5. } 
	
	\begin{figure}[h!]
		\centering
		\includegraphics[width=2.8in]{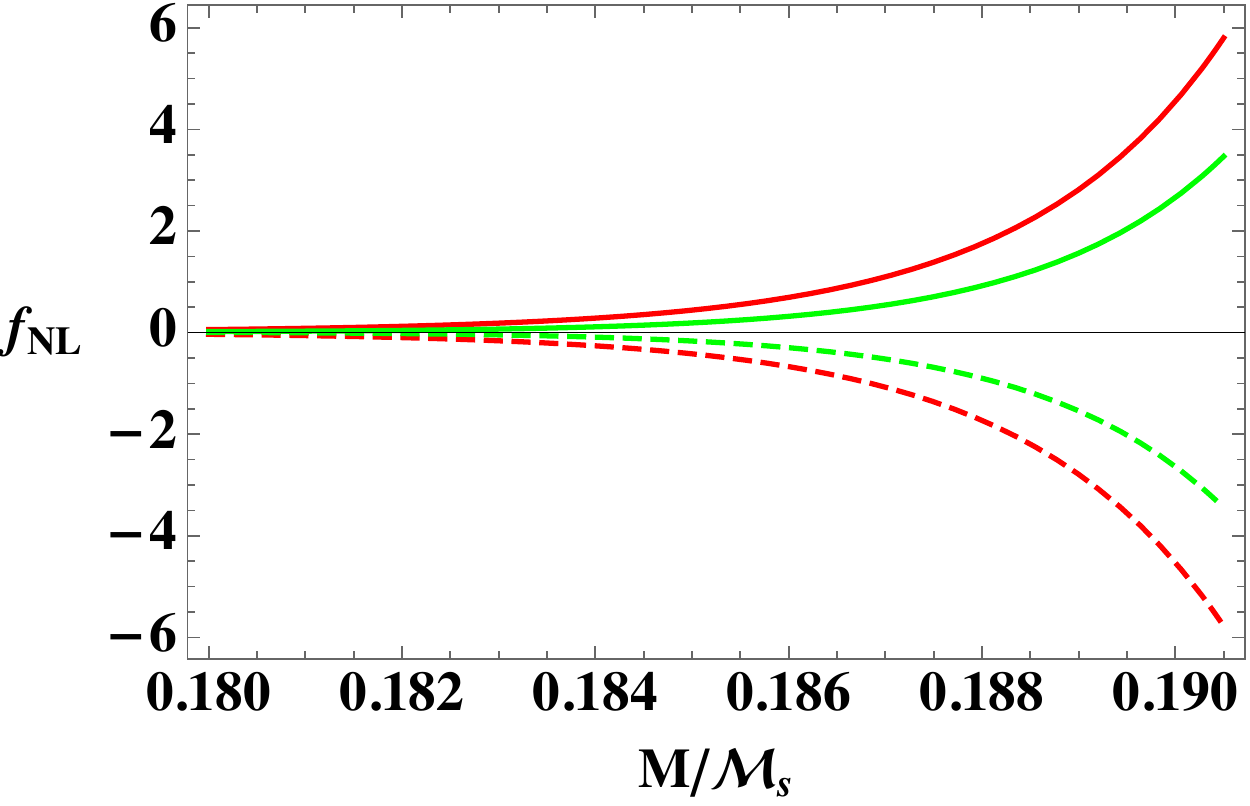}\quad \includegraphics[width=2.8in]{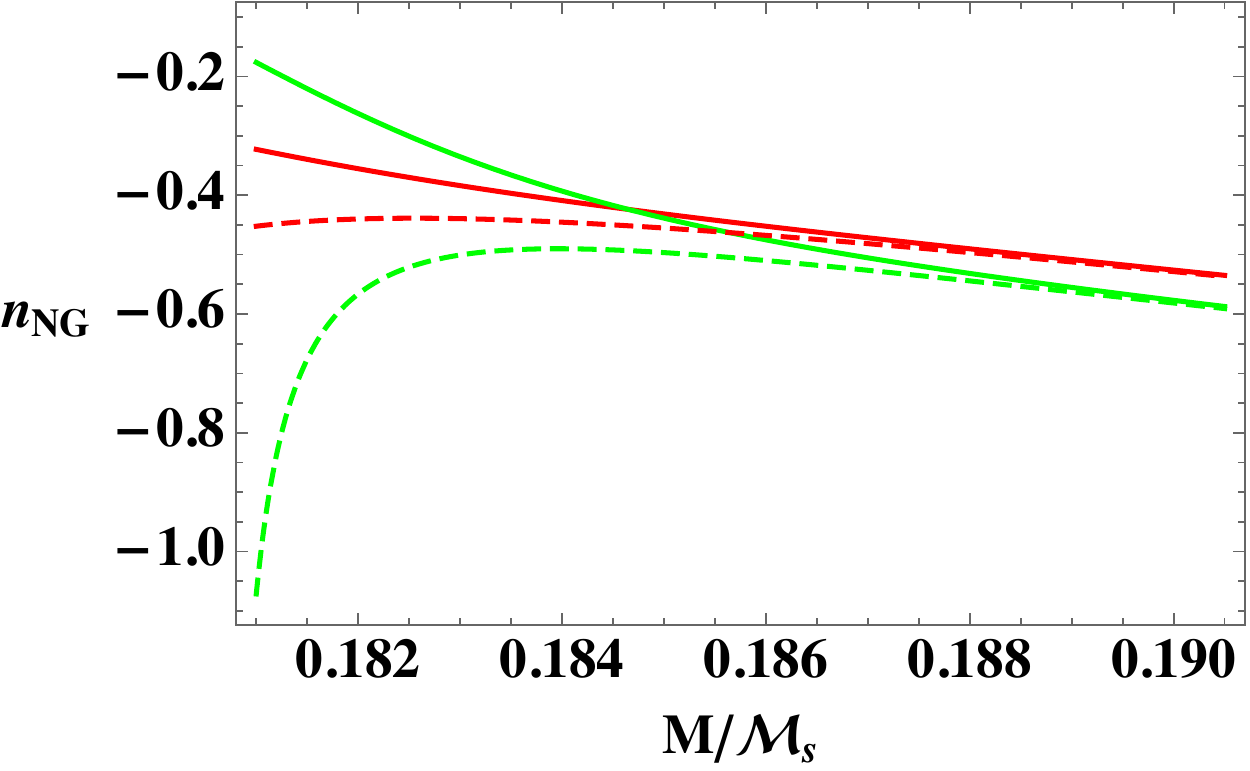}\quad\includegraphics[width=2.8in]{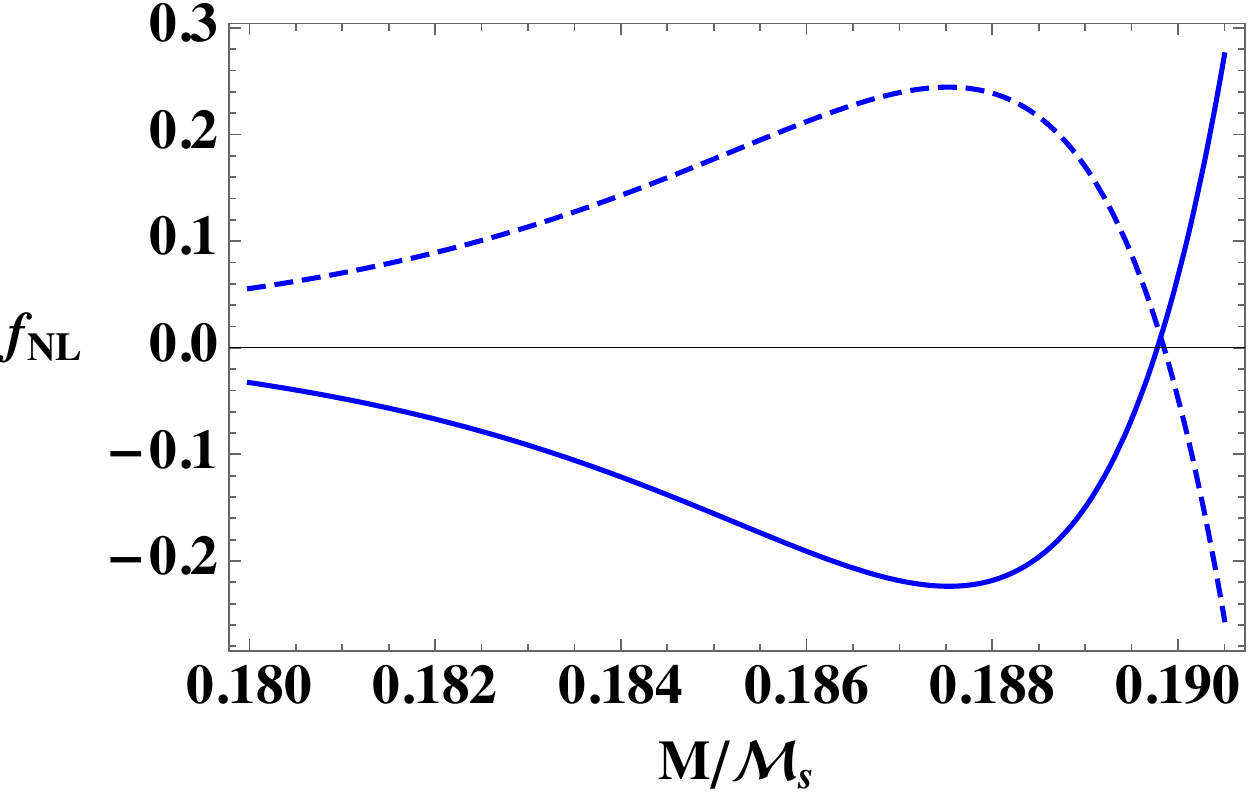}\quad\includegraphics[width=2.8in]{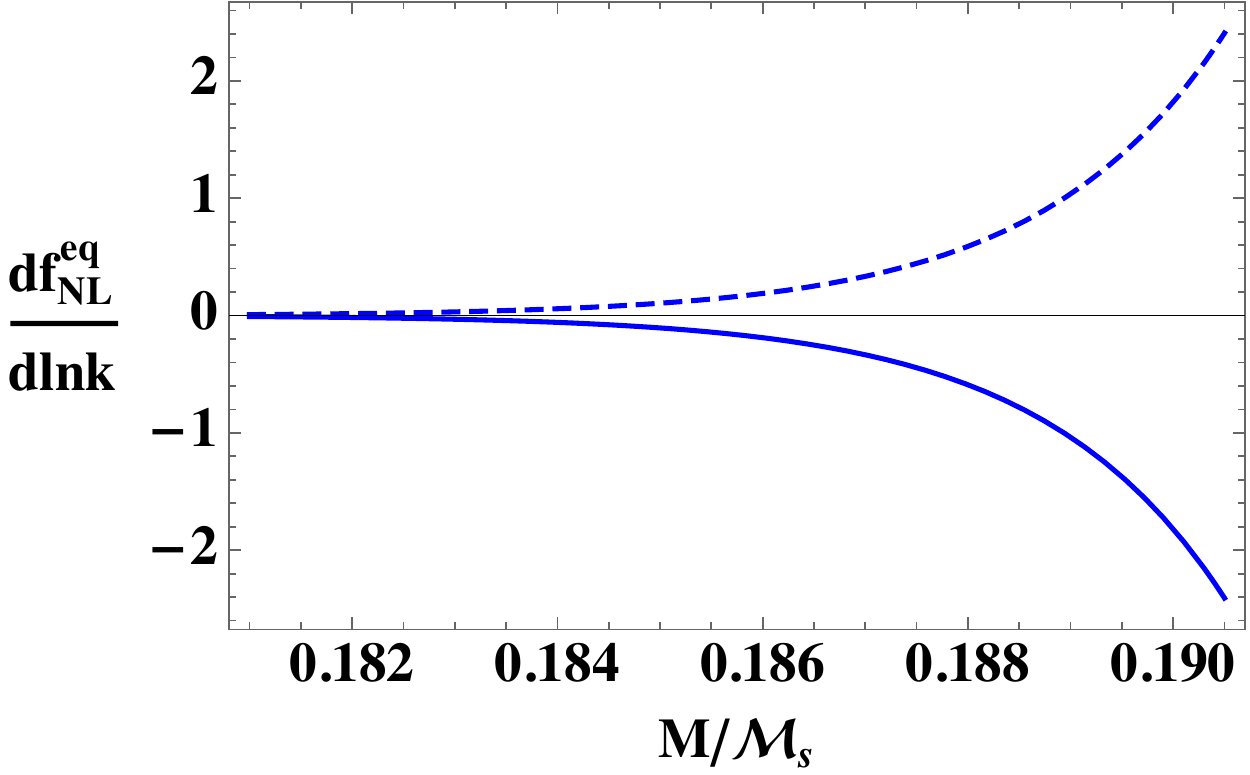}
		\caption{In above plots we present {various limits of $f_{\rm NL}$} and their running $f_{\rm NL},\,n_{\rm NG}$ versus $M/\Mc_s$ for  $\lambda_c>0$ (dashed lines) and $\lambda_c<0$ (full lines) respectively, while red, blue and green lines represent squeezed, equilateral and orthogonal {limits} respectively. For all plots we fix the parameters as $\lambda_c = \pm 2$, $\alpha_{1}=0.2,\,\alpha_{11}= 5, \alpha_{21}= \alpha_{31} = -1$ and $N=55$ corresponding to the pivot scale $k_\ast=a_\ast H_\ast$. }
		\label{R2zsor}
	\end{figure}

	\begin{figure}[h!]
		\centering
		\includegraphics[width=2.8in]{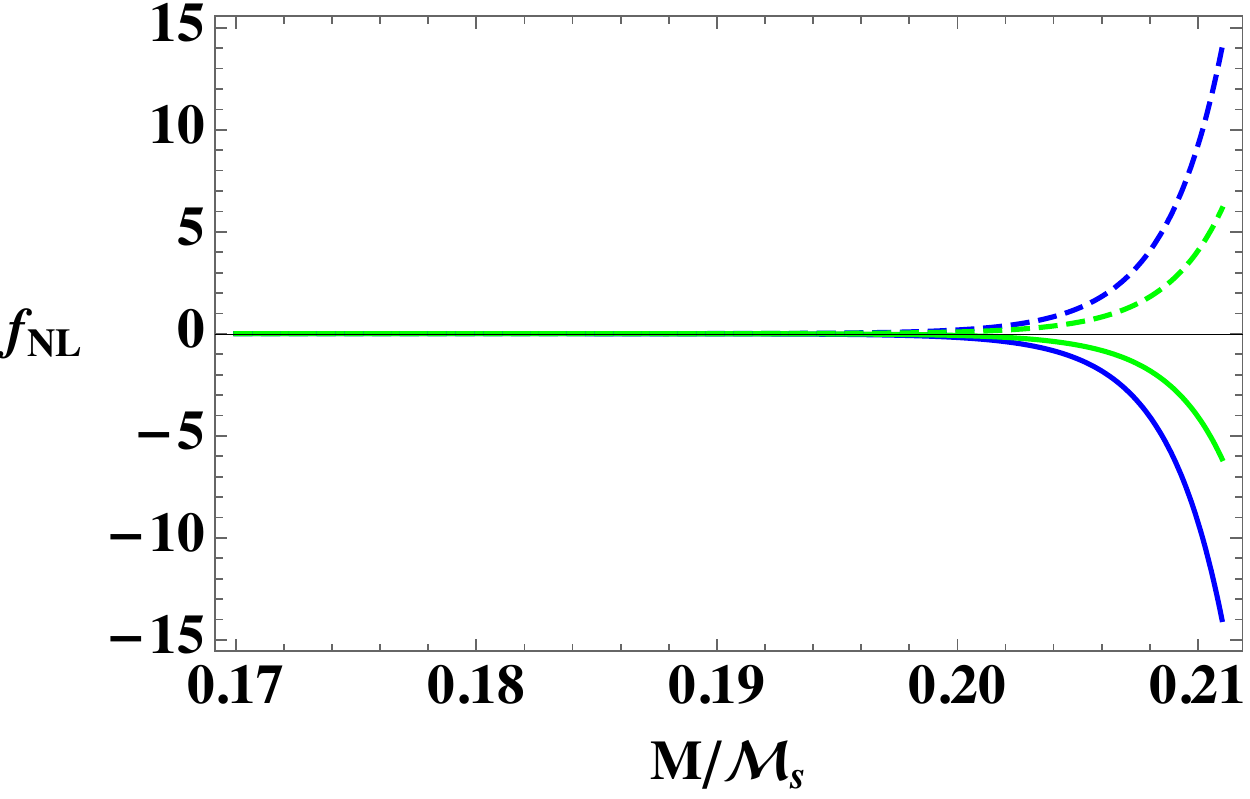}\quad \includegraphics[width=2.8in]
		{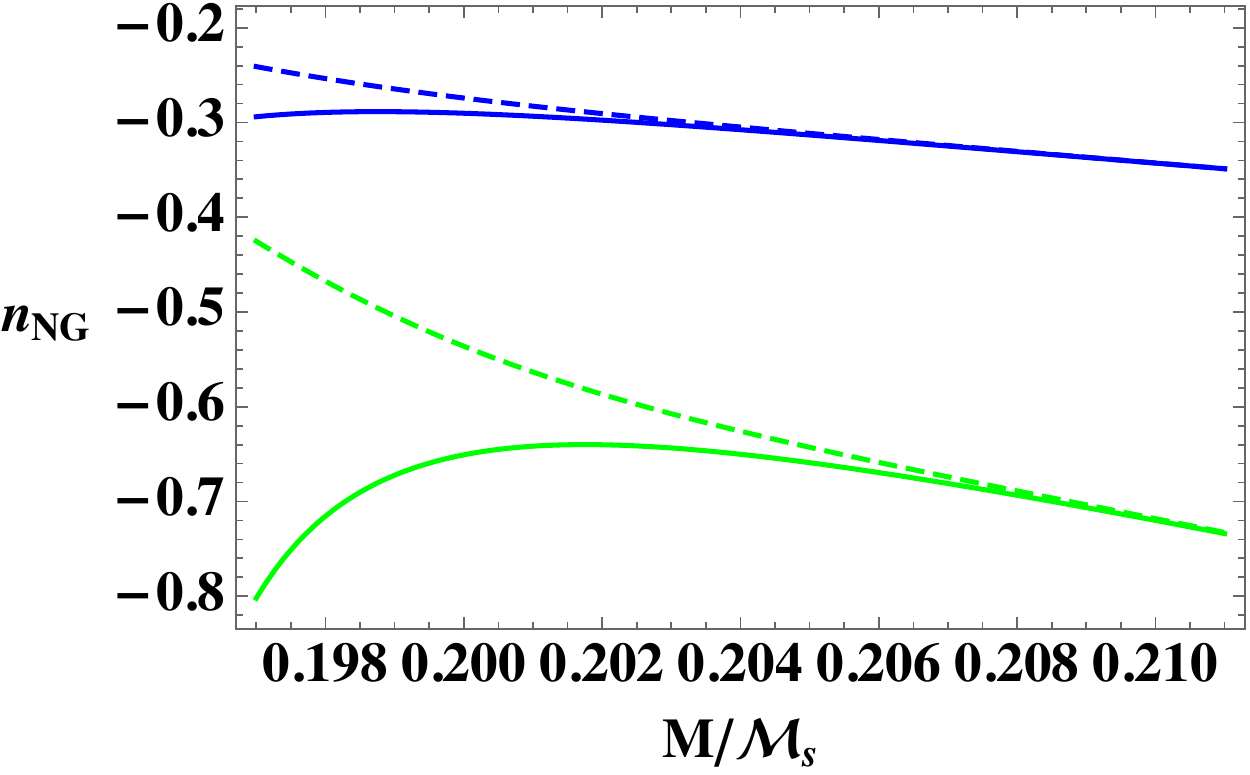}\quad  \includegraphics[width=2.8in]{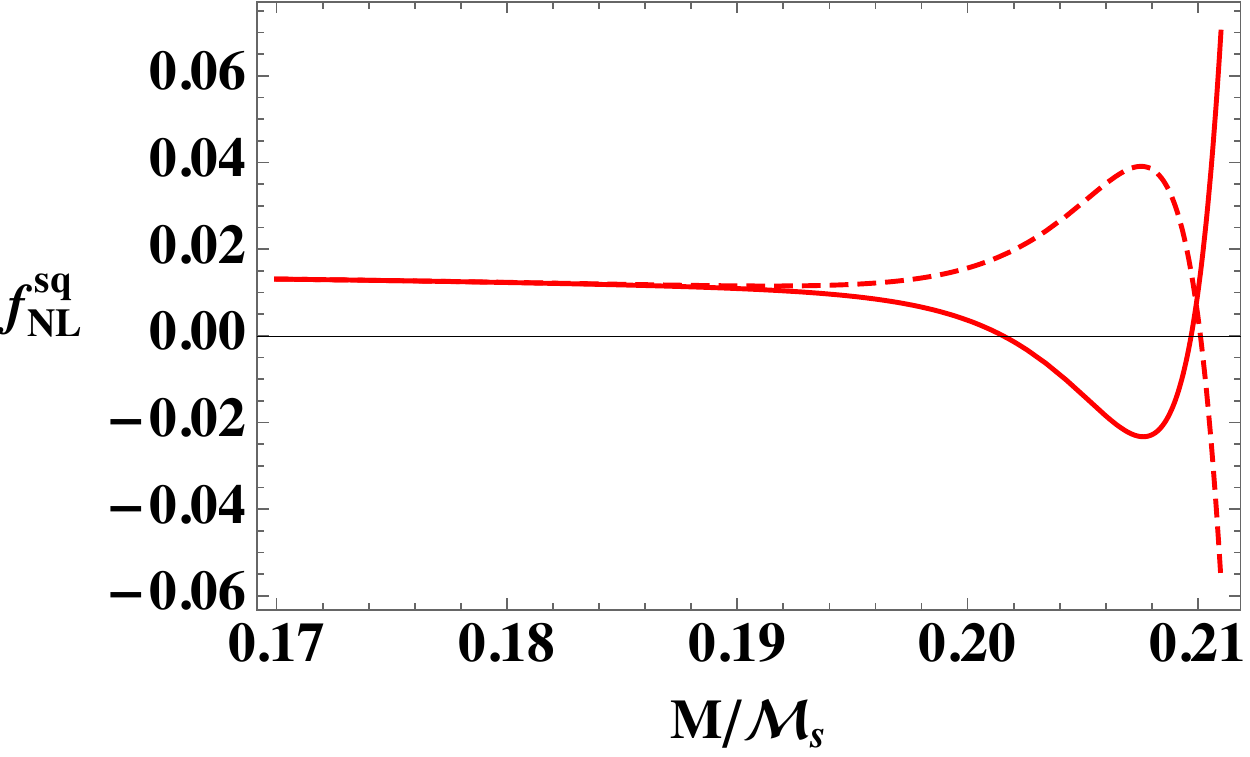}\quad \includegraphics[width=2.8in]{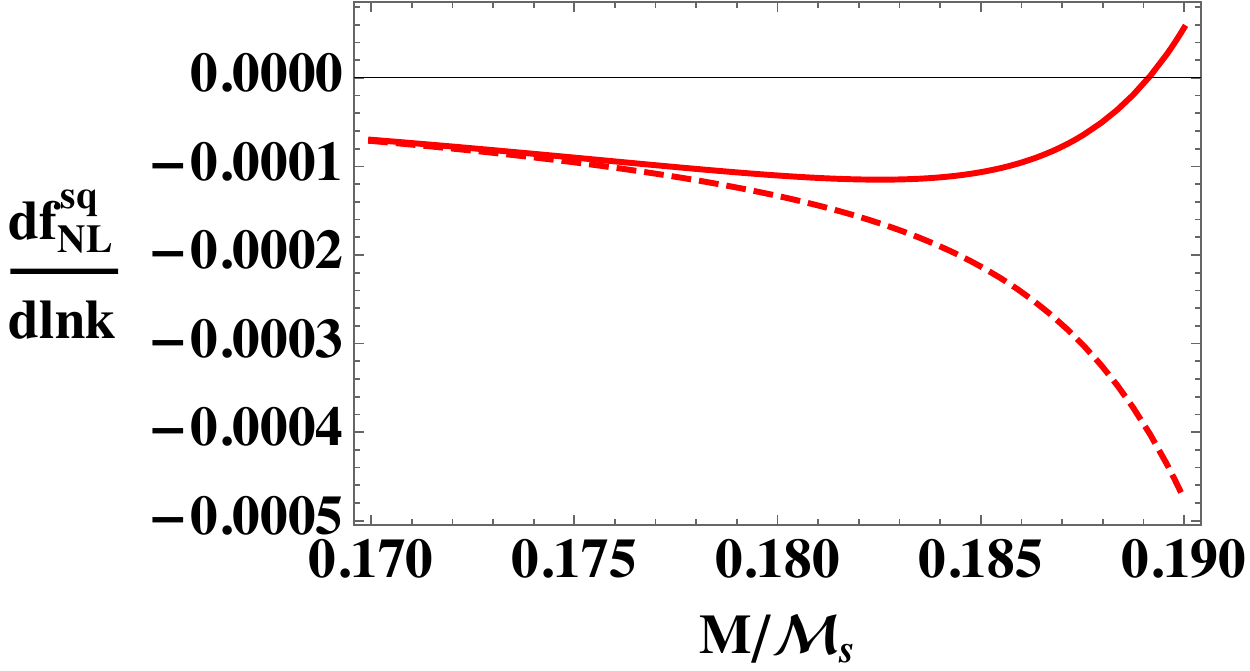}
		\caption{In the above plots we present various limits of $f_{\rm NL}$ and their running $n_{\rm NG}$ versus $M/\Mc_s$. The red, blue and green lines represent squeezed, equilateral and orthogonal {limits} respectively. For all these plots we fix the parameters as $\lambda_c= \pm 2$ (-2 for full lines and +2 for dashed lines), $\alpha_1=0.2,\,\alpha_{11}=3.5,\,\alpha_{21}= \alpha_{31}=-1/4$ and $N=55$ corresponding to the pivot scale $k_\ast=a_\ast H_\ast$. }
		\label{R2zsq}
	\end{figure}

	\item Large PNGs with equal contributions in the squeezed equilateral,  orthogonal limits: see Fig.~\ref{R2zeq}.
	
	\begin{figure}[h!]
		\centering
		\includegraphics[width=2.8in]{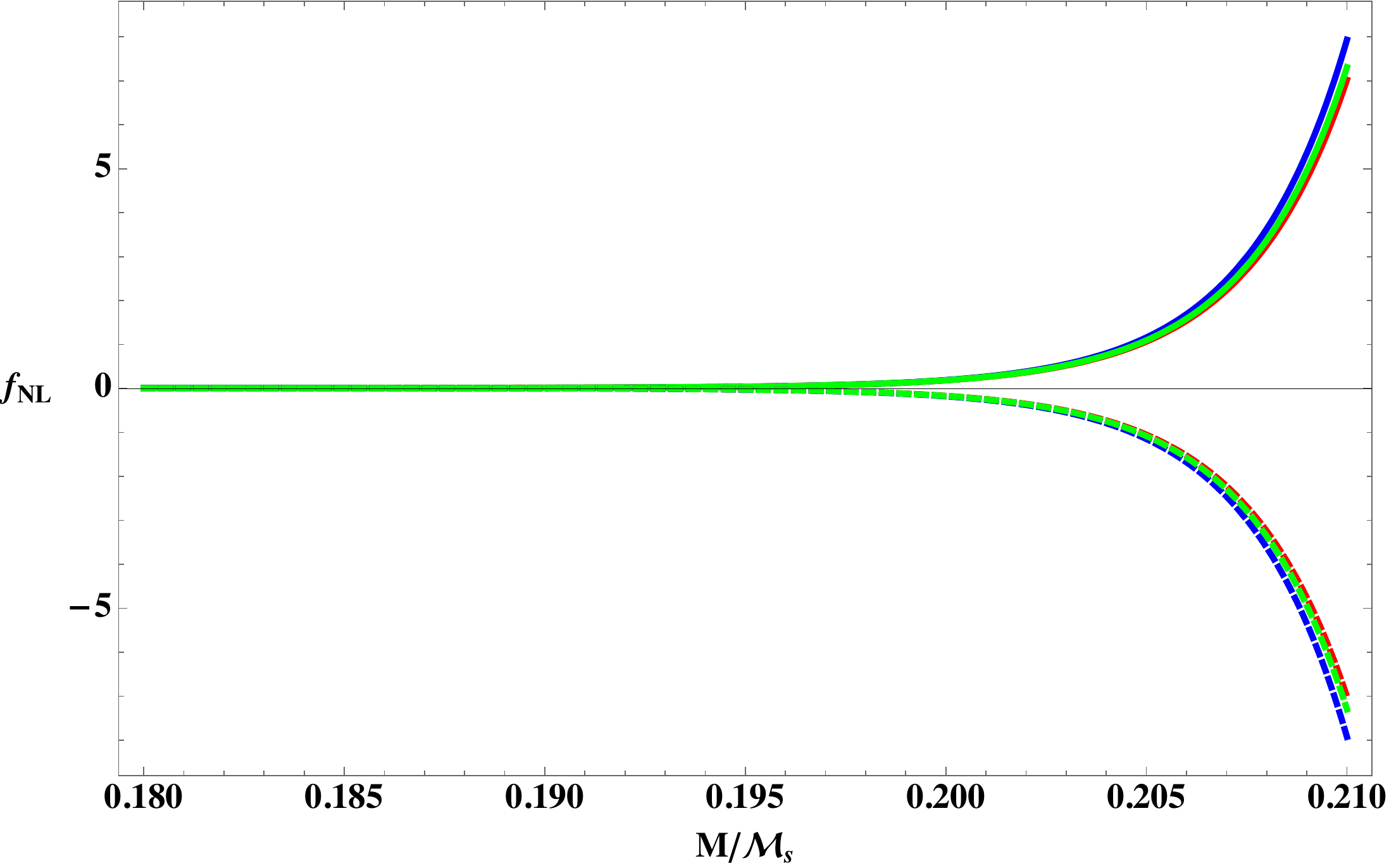}\quad \includegraphics[width=2.8in]{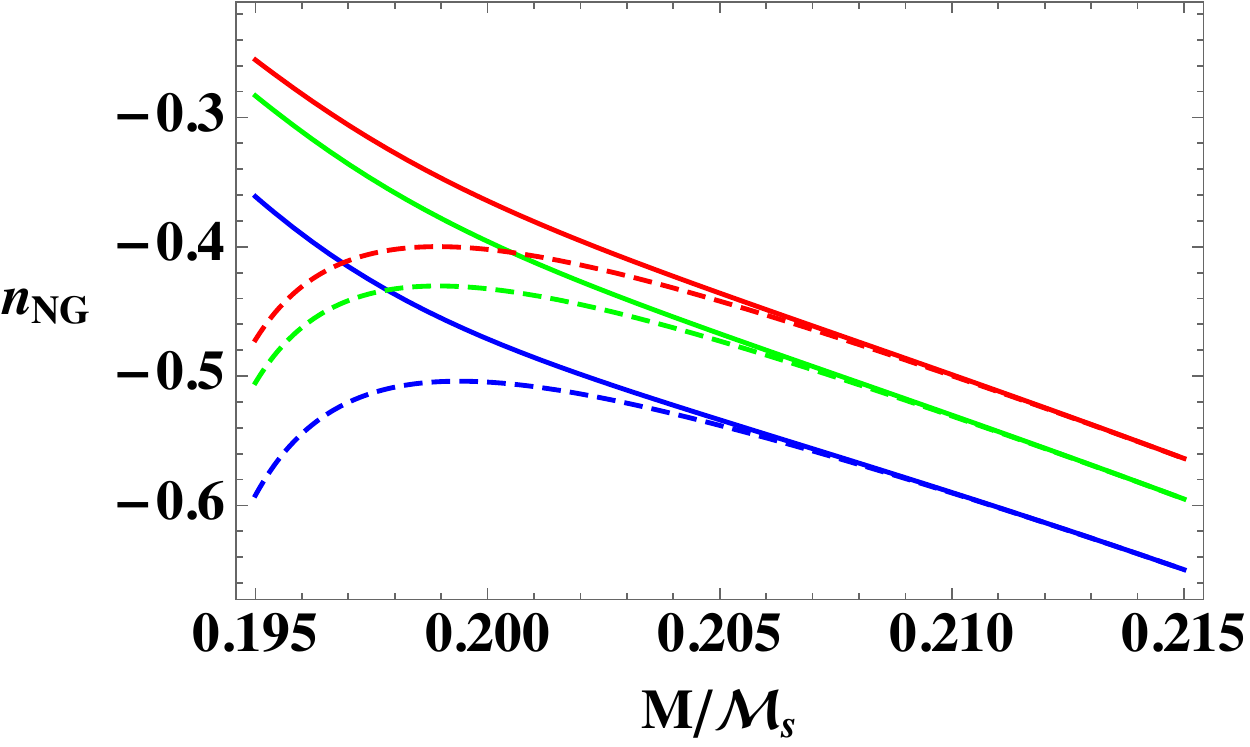}
		\caption{In both plots we present various limits of $f_{\rm NL}$ and their running $n_{NG}$ versus $M/\Mc_s$ for  $\lambda_c>0$ (full lines) and $\lambda_c<0$ (dashed lines) respectively, while red, blue and green lines represent squeezed, equilateral and orthogonal {limits} respectively. For both plots we fix the parameters as $\lambda_c = \pm 2$ (+2 corresponding to full lines and -2 represented by lines) $\alpha_{1}=0.2,\,\alpha_{11}= 3, \alpha_{21}= -3,\,\alpha_{31} = -1$ and $N=55$ corresponding to the pivot scale $k_\ast=a_\ast H_\ast$. }
		\label{R2zeq}
	\end{figure}
	
\end{enumerate}
Finally, we plot the full shape  of $f_{\rm NL}\LF 1,\,\alpha_s,\,\beta_s \RF$  in Fig.~\ref{fig:3Dfnl} as a function of $\LF \alpha_s,\,\beta_s \RF$ defined in \eqref{alsbs}  considering the same parameter values used in Fig.~\ref{R2CFNL}.  {There are several ways one can define the shape of bispectrum \cite{Chen:2010xka}. One definition of the shape function is $S\LF k_1,\,k_2,\,k_3 \RF = -\frac{3}{10} \LF \sum_i k_i^3\RF \frac{f_{NL}\LF k_1,\,k_2,\,k_3 \RF }{k_1k_2k_3}$ \cite{Chen:2010xka}. In this paper, we stick to the shape function defined by the reduced bispectrum $f_{\rm NL}\LF k_1,\,k_2,\,k_3 \RF$ \cite{Kenton:2015lxa} to represent the 3D plot in Fig.~\ref{fig:3Dfnl}.} 

{Following the discussion we presented towards the end of Sec.~\ref{sec:NGs} (where we compared the interactions of curvature perturbation in our non-local theory with analogous interactions in local EFTs \cite{Chen:2010xka,Chen:2006xjb,Chen:2006nt})  we can deduce that even though shapes of PNGs in non-local $R^2$-like inflation appears to be a mixture of so-called squeezed, equilateral and orthogonal shape templates \cite{Chen:2010xka}, we obtain here large non-trivial running of PNGs due to the strong scale dependencies due to non-localities which we discussed in Sec.~\ref{sec:NGs}. Therefore, the running of PNGs plays an important role in distinguishing PNGs in non-local $R^2$-like inflation against local multifield non-canonical scalar fields inflation \cite{Chen:2010xka}.  {This is similar to the running of tensor spectral index that was obtained in the our earlier work \cite{Koshelev:2022olc} which acts as a distinguishable feature of non-local $R^2$-like inflation 
against local EFTs at the level of the two-point tensor correlation function.}
However, we defer the detailed analysis of shape of PNGs for the full parameter space for future investigations. }

\begin{figure}
	\centering
	\includegraphics{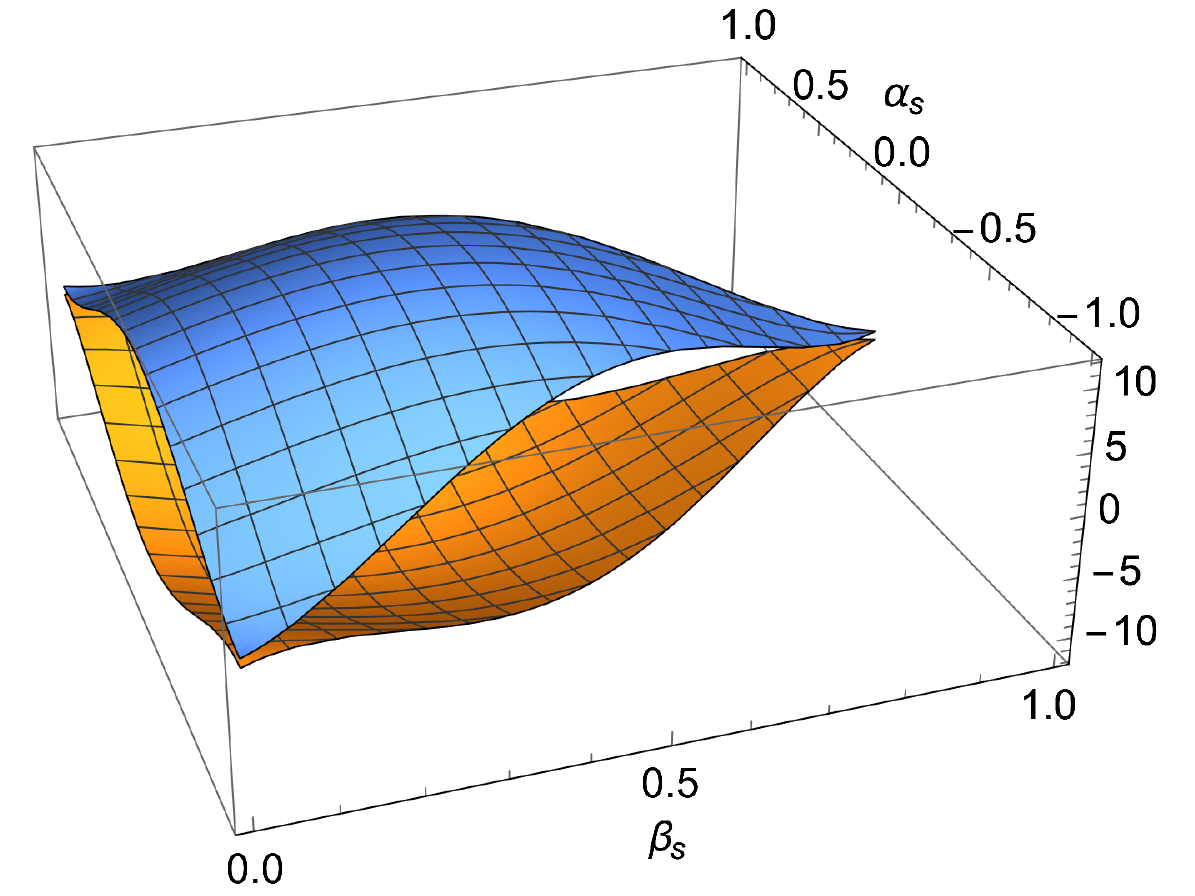}
	\caption{Here we plot $f_{\rm NL}\LF 1,\,\alpha_s,\,\beta_s \RF$ setting $K=1$. The values of parameters for this figure are the same as those for the Fig.~\ref{R2CFNL}. In this figure, the upper plane in blue shape and lower orange shape represent $\lambda_c>0$ and $\lambda_c<0$ respectively.   }
	\label{fig:3Dfnl}
\end{figure}

\section{Distinguishing generalized non-local $R^2$-like inflation against EFT of scalar field inflation}
\label{sec:eftcom}

In a broader view of developments in inflationary cosmology, we can notice that there are several ways one can construct a viable inflationary phase. Due to the lack of concrete understanding of UV-complete physics, numerous inflationary models were proposed in the past decades \cite{Martin:2013tda} taking multiple approaches with different motivations towards UV-completion. In this respect, inflationary cosmology is hugely studied in the context of EFT involving single and multiple scalar fields of different nature \cite{Cheung:2007st,Senatore:2010wk,Kobayashi:2011nu}. In view of inflationary observables related to 3-point correlations, 
%herewith the following classification of EFTs 
we shall discuss how to distinguish EFT of several classes of inflationary models against the generalized non-local $R^2$-like inflation we have studied in this paper. Note that distinguishing the generalized non-local $R^2$-like inflation against the EFT of inflation was extensively discussed in \cite{Koshelev:2022olc}.

\textbf{Canonical single field inflation.} This is the simplest framework of understanding inflation driven by a scalar field with the canonical kinetic term and a suitable potential. The successful models of this class are $R^2$ and Higgs inflation after the conformal transformation to the Einstein frame \cite{Akrami:2018odb}. 
Independent of what is the choice of potential, this class of models always satisfy the tensor consistency relation and the Maldacena consistency relation. In our non-local gravity \eqref{NAID} these two consistency relations are violated (see \cite{Koshelev:2022olc} and our result \eqref{fnlseo}). However, it has been widely understood that such models cannot be really single field ones because any attempt of embedding the models in a UV-complete framework introduce more degrees of freedom  which eventually violate the consistency relation (although the new particles which appear in this way could be heavier than the inflaton) \cite{Linde:2014nna,Noumi:2012vr,Arkani-Hamed:2015bza,Lee:2016vti}.   

\textbf{Non-canonical single field inflation.}
Non-canonical scalar field ($\phi$) is described by 
\begin{equation}
	S_{\rm NC} = \int d^4x\sqrt{-g}\Bigg[ \frac{M_p^2}{2}\,R+P\LF-\frac{1}{2}\LF \pd \phi \RF^2,\,\phi\RF \Bigg]
\end{equation}
where $P(-\frac{1}{2}\LF \pd \phi \RF^2,\,\phi) $ is a generic function of the canonical kinetic term and the field $\phi$. This is a general action for a scalar field whose perturbations propagate with a sound speed $0\leq c_s\leq 1$. Because of this, non-canonical models of inflation can predict a peak in the equilateral {limit}, the tensor consistency relation is violated while tensor tilt remains unaffected by $c_s$:
\begin{equation}
	f^{\rm sq}_{\rm NL} \sim \frac{1}{c_s^2},\quad r= 16c_s\epsilon_E,\quad n_t = -2\epsilon_E
	\label{ncmodel}
\end{equation}
where the slow-roll parameter with subscript $\epsilon_{E}$ denotes the quantity in the canonical scalar field framework of inflation.
With the Planck constraints \eqref{boundsfnl}, we get $c_s\gtrsim 0.02$ \cite{Akrami:2019izv}. There are two ways we can distinguish our model from a non-canonical scalar one. One way is that in our case we modify tensor tilt and it becomes scale dependent (see \cite{Koshelev:2022olc}). Another way is provided by the shape of PNGs, in the non-canonical model it is determined by the following interaction terms of curvature perturbation  
\begin{equation}
	\begin{aligned}
		\delta^{(3)}S_{\rm NC}	=	\int d^3 x \, d\tau \; a^2	\Bigg\{ &
		\frac{g_1}{a} \zeta'^3
		+ g_2 \zeta \zeta'^2
		+ g_3 \zeta (\partial \zeta)^2
		+ g_4 \zeta' \partial_j \zeta \partial_j \partial^{-2} \zeta'
	\\ &	+ g_5 \partial^2 \zeta
		(\partial_j \partial^{-2} \zeta')
		(\partial_j \partial^{-2} \zeta')
		\Bigg\} 
	\end{aligned}
	\label{ncint}
\end{equation}
where $\zeta$ is the curvature perturbation ($\zeta \approx -\Rc$ at super-Hubble scales) and $g_i$ are functions of slow-roll parameters \cite{Burrage:2011hd}. If we compare these interaction terms with those we obtained in our theory \eqref{3rda}, we can deduce that the {large PNGs are of a possibility}, and if the future observations will be powerful enough, we can distinguish our non-local gravity from non-canonical single field models of inflation. \\

\textbf{Multifield inflation.} Multifield inflation is basically an inflationary scenario driven by multiple scalar fields that leads to curvature and isocurvature perturbations during inflation \cite{Wands:2007bd}. Therefore several classes of multifield inflation were studied in literature \cite{Byrnes:2014pja,Vennin:2015vfa}. Due to the presence of isocurvature modes, curvature perturbation, if defined as being tangent to the background trajectory of inflaton fields in the field space, is generically not time-independent on super-Hubble scales\footnote{Note, however, that there always exists another mode (a particular solution) of scalar perturbations in this case which is (approximately) time-independent in the super-Hubble regime~\cite{Starobinsky:1994mh,Starobinsky:2001xq} though it is not tangent to the inflaton background trajectory in the field space generically. Its existence can be traced to the fact that the FLRW scale factor is defined up to an arbitrary multiplier in the absence of spatial curvature, see~\cite{Starobinsky:2001xq} for more discussion.} that leads to violation of the Maldacena consistency relation due to the classical evolution of fluctuations on such scales \cite{Vernizzi:2006ve,Wands:2007bd}. As a result of this, the squeezed limit $f^{\rm sq}_{\rm NL}$ gives detectable level of non-Gaussianities. In our case we obtain detectable level of $f^{\rm sq}_{\rm NL}$ due to the non-local interactions of short and long wavelength modes which are quantum mechanical in nature (See Sec.~\ref{sec:NGs} and Appendix~\ref{app:violmc}). Therefore, we generate $f^{\rm sq}_{\rm NL}$ in our model in a totally different way compared to multifield inflation. Furthermore, in the context of multifield inflation $f^{\rm sq}_{\rm NL}$ is almost scale independent \cite{Wands:2007bd}. Moreover, multifield inflation with non-canonical scalar field predicts {several types} of PNGs {(with large signatures in squeezed, equilateral and orthogonal limits)} \cite{Gao:2008dt}. In our case, we obtain {various class of PNGs} in a single field geometric construction of inflation.  Particularly, in our case, we generate scale dependent PNGs with detectable running (see Sec.~\ref{sec:NGs} and Sec.~\ref{sec:NGRUN}).  Furthermore, tensor consistency relation is violated in multifield models of inflation due to the classical evolution of curvature perturbation in the presence of additional isocurvature fields~\cite{Polarski:1995zn}. But this feature can be distinguished from our case because we modify the tensor power spectrum that leads to corrections to the tensor spectral index $n_t$ and its running which are derived in \cite{Koshelev:2022olc}. 

\textbf{EFT of single-field inflation.} 

EFT of single field inflation (EFT-SI) \cite{Cheung:2007st,Weinberg:2008hq} is constructed to unify several single field models of inflation and aims to capture the signatures of UV-completion through the so-called EFT parameters that determine 2-point and 3-point inflationary correlations. It was elaborated in \cite{Koshelev:2022olc} that the physics of two-point correlations in our generalized non-local gravity inflation with \eqref{NAID} cannot fall into the category of EFT-SI.  The key EFT-SI parameters that affect both 2-point and 3-point correlations are the sound speeds $\LF c_s,\,c_t \RF$  of scalar (curvature) and tensor fluctuations which are slowly varying functions of time. %that are bounded as $0\leq \LF c_s,\,c_t \RF\leq 1$. 
In the EFT model considered in \cite{Creminelli:2014wna}, $c_s=1, c_t\not= 1$ in the original (disformal) frame, while $c_{sE}=\frac{1}{c_t}, c_{tE}=1$ in the Einstein frame obtained  
%These two parameters are related to each other $c_s=\frac{1}{c_t}$ that is derived 
through the disformal transformation which changes the light cone. This leads to relation between scalar and tensor PNGs. In contrast to this, in our case of non-local gravity inflation we have both sounds speeds being unity and the scalar PNGs in our case are determined by the parameter space of the part of our full action \eqref{NAID} excluding the terms of the Weyl tensor. Therefore, we do get additional contributions which depend on additional parameter space for tensor PNGs that is the subject of our future investigation \cite{KKS2}. This means that our geometric construction of inflation is indeed clearly distinguishable from that of the EFT-SI \cite{Cheung:2007st,Weinberg:2008hq}. Focusing on the scalar PNGs in particular, the 3rd order action of EFT-SI prescribes 
\begin{equation}
	\begin{aligned}
		\delta_{(s)}^{(3)}S_{\rm EFT-SI} = \int d^4x\sqrt{-g} \Bigg[\frac{M_p^2}{c_s^2}\epsilon_E \zeta^2\Big( \dot{\zeta}^2-c_s^2\frac{(\pd_i\zeta)^2}{a^2} \Big)& -M_p^2 \epsilon_E \LF 1-\frac{1}{c_s^2} \RF \dot{\zeta}\frac{(\pd_i\zeta)^2}{a^2}\\ &+\Bigg(\frac{M_p^2}{H}\epsilon_E \LF 1-\frac{1}{c_s^2} \RF +\frac{4}{3}\frac{M_3^4}{H^3}\Bigg)\dot{\zeta}^3\Bigg]\, .
	\end{aligned}
	\label{EFTSI}
\end{equation}
The above action is derived in the limit then $\epsilon_E \ll c_s^2$ and it 
contains two interactions $\dot{\zeta}^3$ and $\dot{\zeta}\LF\pd_i\zeta\RF^2$ associated with two parameters $\LF c_s,\, M_3 \RF$. These two interactions can be projected into equilateral and orthogonal shapes and the recent Planck data constrained them \cite{Akrami:2019izv}. Comparing the interactions in \eqref{EFTSI} with our case \eqref{3rda}, we can easily deduce that non-local gravity does contain more interactions and scale dependencies beyond the standard EFT-SI. Therefore, it would be of great interest to observationally probe \eqref{3rda} in the future CMB and LSS observations \cite{Meerburg:2019qqi}.  

\textbf{EFTs of quasi-single field inflation (aka cosmological collider physics).} This is a paradigm where there is one light scalar degree of freedom driving inflation interacting with heavy modes whose mass is of the order of the Hubble scale during inflation \cite{Chen:2010xka,Noumi:2012vr,Lee:2016vti}. This framework of inflation is a natural set up in several UV-complete theories such as supergravity \cite{Bartolo:2018hjc,Hetz:2016ics,Aoki:2020zbj,Arkani-Hamed:2015bza}. In this class of inflationary models equilateral, orthogonal and squeezed limit PNGs can be generated due to turns in the inflaton trajectory in the field space (that occur due to the presence of additional heavy fields). 
The crucial test of quasi-single field inflation is the shape of PNG that is scale dependent in the squeezed limit as follows \cite{Chen:2015lza}
\begin{equation}
	f^{\rm sq}_{\rm NL}\LF k_{\rm long},\,k_{\rm short},\,k_{\rm short}\RF  \propto \LF \frac{k_{\rm long}}{k_{\rm short}} \RF^{\nu_s} P_{\bar{s}}\LF \cos\theta \RF\ , 
	\label{quasising}
\end{equation}
where $P_{\bar{s}}\LF \cos\theta \RF$ is the function depending on $\bar{s}$ which is the spin of the heavy particle, $\theta$ is the angle between long and short wavelenghth wave vectors and 
\begin{equation}
	\nu_s = \frac{3}{2}\pm i\sqrt{\frac{m^2}{H^2}-\frac{9}{4}}
\end{equation}
with $m$ being the mass of the heavy particle. Depending on the mass of the heavy particle, we get different scale dependent features in the PNGs. This is a different scale dependence compared to what we have obtained in \eqref{fnlseo} where $f_{\rm NL}$ has scale dependence on the overall wave number rather than on the short and long wavelength modes in \eqref{quasising}. This is the distinguishable feature of non-local theories from EFT of quasi-single field inflation. 

\section{Outlook}

In this paper, we have computed and analyzed scalar PNGs which can be generated in the generalized non-local $R^2$-like inflation and we provided detailed discussion about how our gravity framework of inflation goes beyond the several EFTs of inflation. Quantitative description of PNGs and their running is obtained, too. However, through scalar PNGs we only explored part of the action \eqref{NAID} which contributes to $f_{\rm NL}$ in the leading order. It is important to further understand tensor PNGs and cross-correlations that is the subject of our future investigation \cite{KKS2}. Another direction to go forward is to compute 4-point correlations in the generalized non-local gravity inflation and check if the Suyama-Yamaguchi consistency relation \cite{Suyama:2007bg} gets violated in non-local theories. Furthermore, it is interesting to explore the PNGs which can emerge if we couple matter degrees of freedom non-minimally and non-locally to gravity \cite{Draper:2020knh}. Beyond these studies, it is theoretically very important to study formfactors of our gravity action \eqref{NAID} from a more fundamental point of view that can provide us with more robust and precise predictions.

\label{sec:Conc}

\acknowledgments
AK is supported by FCT Portugal investigator project IF/01607/2015. This research work was supported by grants  UID/MAT/00212/2019, COST Action CA15117 (CANTATA). KSK acknowledges the support from JSPS and KAKENHI Grant-in-Aid for Scientific Research No. JP20F20320 and No. JP21H00069. 
KSK would like to thank the Royal Society for the Newton International Fellowship.
AAS was supported by the RSF grant 21-12-00130.  We would like to thank L.~Buoninfante, A.~De~Felice, T.~Noumi, A.~Tokareva, T.~Suyama and M.~Yamaguchi for very useful discussions. 

\appendix

\section{Computation of non-Gaussianities}
\label{App3NG}

In this section we compute the PNG contributions from the term which is cubic in Ricci scalar in \eqref{NAID}. 
\begin{equation}
	\begin{aligned}
		\delta^{(3)}S_{R^3} =\, & \frac{f_0\lambda_c}{2\Mc_s^2} \int d^4x \sqrt{-\bar{g}}  \Bigg[ \Lc_1\LF \bar{\square}_s \RF \delta R \Lc_2\LF \bar{\square}_s \RF \delta R \Lc_2\LF \bar{\square}_s \RF \delta R \\ & + \delta \Lc_1\LF \square_{s} \RF \bar{ R} \delta \Lc_2\LF \square_{s} \RF \bar{ R} \delta \Lc_3\LF \square_{s} \RF \bar{ R}\Bigg] 
	\end{aligned}
	\label{delSR3}
\end{equation}
We can notice that \eqref{delSR3} does not contain any terms involving second variations of Ricci scalar and the 3rd variations of $\sqrt{-g}$ such as
\begin{equation}
	\begin{aligned}
		\bullet\,\,& \Lc_i\LF \square_s \RF \delta^{(2)}R   \Lc_j\LF \square_s \RF \bar{R}\Lc_k\LF \square_s \RF \delta R\\ 
		\bullet\,\,&  \Lc_i\LF \square_s \RF \delta^{(2)}R   \Lc_j\LF \square_s \RF \bar{R}\Lc_k\LF \square_s \RF \delta \bar{R}\\ 
		\bullet\,\,&
		\delta^{(2)}\sqrt{-g}  \Lc_i\LF \square_s \RF \bar{R} \Lc_j\LF \square_s \RF \bar{R} \Lc_k\LF \square_s \RF \bar{R}  
	\end{aligned}
\end{equation}
since any such terms be multiplied by $\Lc_{i}\LF \bar{	\square}_s \RF\bar{	R}$ which become zero after imposing the on-shell condition \eqref{entic}.
Further simplifying \eqref{delSR3} using the following formula \cite{Koshelev:2020foq}
\begin{equation}
	\delta \Lc_i\LF \square_s \RF = \sum_n L_{in} \sum_{a+b=n-1} \bar{\square}_s^a \delta\square_s \bar{\square}_s^b\,, 
\end{equation}
we obtain 
\begin{equation}
	\begin{aligned}
		\delta^{(3)}_{(s)}S^{\rm Non-local}_{R^3}  = &\, 
		\frac{f_0\lambda_c}{2\Mc_s^2} \int d^4x \sqrt{-\bar{g}}  \Bigg[  \Bigg(\Lc_1\LF \bar{\square}_s \RF \delta R+\frac{\Lc_1\LF \bar{\square}_{s} \RF}{\bar{\square}_{s}-\frac{M^2}{\Mc_{s}^2}} \delta\square_{s}\bar{ R}\Bigg) \\ &  \Bigg(\Lc_2\LF \bar{\square}_s \RF \delta R+\frac{\Lc_2\LF \bar{\square}_{s} \RF}{\square_{s}-\frac{M^2}{\Mc_{s}^2}} \delta\bar{\square}_{s}\bar{ R}\Bigg) \Bigg(\Lc_3\LF \bar{\square}_s \RF \delta R+\frac{\Lc_3\LF \bar{\square}_{s} \RF}{\bar{\square}_{s}-\frac{M^2}{\Mc_{s}^2}} \delta\square_{s}\bar{ R}\Bigg) \Bigg] \,\\ 
		= & -\frac{f_0\lambda_c\epsilon^3}{2\Mc_{s}^2}\int d^4x \sqrt{-\bar{g}}  \Bigg\{  \Bigg[ \frac{64}{a^6}\Hc^3  \Bigg(8\epsilon^3 T_{\rm NL}^4  + 4\epsilon^2 T_{\rm NL}^3 \\ & 2\epsilon T_{\rm NL}^2+T_{\rm NL}^1\Bigg) \Rc^\prime\Rc^\prime\Rc^\prime + \frac{64}{a^4}\bar{R}_{\rm dS} \Hc^2 \LF  4\epsilon^2 T_{\rm NL}^3+ 2\epsilon T_{\rm NL}^2 + T_{\rm NL}^1\RF \Rc\Rc^\prime\Rc^\prime \\ &  +\frac{128}{a^2}\bar{R}_{\rm dS}^2 \Hc\LF  \epsilon T_{\rm NL}^2+ T_{\rm NL}^1 \RF \Rc\Rc\Rc^\prime+ 64 \bar{R}_{\rm dS}^3 T_{\rm NL}^1\Rc^3  \Bigg]\Bigg\}\,. 
	\end{aligned}
	\label{3rdOac}
\end{equation}
where $\Hc= aH$ is the Hubble factor in conformal time. Here we have used the slow-roll approximation in the next to leading order in $\epsilon$ and made a substitution $\bar{	\square}_{\rm dS}\Rc = M^2\Rc$ which solution in the quasi-dS approximation is 
\begin{equation}
	\Rc \approx -\frac{1}{2\epsilon}\frac{1}{\sqrt{3f_0\bar{	R}_{\rm dS}}} \frac{H}{\sqrt{2k^3}} \LF 1+ik\tau \RF e^{-ik\tau}\,. 
	\label{rcsol}
\end{equation}
This follows from the computations in \cite{Koshelev:2022olc}.  We have also used the following relations which were derived using quasi-dS approximation (see Eqs (B.5) and (B.7) in \cite{Koshelev:2020foq}): 
\begin{equation}
	\begin{aligned}
		\Oc\LF \bar{\square}_s \RF \delta R  \approx &\, \Oc\LF \frac{2M^2}{\Mc_s^2} \RF  2\bar{R}\Psi + \frac{2\bar{R}}{M^2}H\epsilon \LT \Oc\LF\frac{2M^2}{\Mc_{s}^2}\RF-\Oc(0)\RT \dot{\Psi} \\& +16H\epsilon\LT \Oc\LF \frac{M^2}{\Mc_s^2}+\frac{\bar{R}}{4\Mc_{s}^2} \RF -\Oc(0)\RT \dot{\Psi}\,\\
		\Oc\LF \square_s \RF\delta\square_s \bar{R} & \approx \frac{2M^2}{\Mc_{s}^2} \Oc\LF \frac{2M^2}{\Mc_{s}^2} \RF\bar{R}\Psi - \frac{2\bar{R}H\epsilon}{\Mc_{s}^2} \Oc\LF \frac{M^2}{\Mc_s^2}+\frac{\bar{R}}{4\Mc_{s}^2} \RF\dot{ \Psi}
	\end{aligned}
	\label{listzfz2}
\end{equation}
where $\Oc$ denotes an analytic function of d'Alembertian $\square$. 
In \eqref{3rdOac} we carefully kept all terms up to the  leading order in $\epsilon\approx \frac{1}{2N}$ and neglected higher order contributions of $O\LF \frac{1}{N^2} \RF$ treating $\epsilon\approx \const$. In our computation we re-sum contribution of all the infinite derivative terms. 
In the second line of \eqref{3rdOac}, we made simplifications using \eqref{Commtr}. Since our case is single field inflation, perturbed modes deep inside the Hubble radius during inflation are evaluated at the Hubble radius exit since curvature perturbations remains constant on super-Hubble scales $k\ll aH$. So the 3-point correlations of $\Rc$ do not evolve after that. Therefore, we evaluate the 3-point correlations at the moment when
$\tau_e\sim -\frac{1}{k}$ which is a conformal time after few e-foldings of the Hubble radius exit \cite{Pajer:2016ieg}.

Here
\begin{equation}
	\Tc_{\rm NL}  = \frac{\bar{R}_{\rm dS}}{M_p^2}\epsilon^3   \LT\Fc_{R}\LF \frac{M^2}{\Mc_{s}^2} +\frac{\bar{ R}_{\rm dS}}{4\Mc_{s}^2}\RF - \Fc_{1} \RT  \approx \frac{\bar{R}_{\rm dS}}{M_p^2}\epsilon^3 \LT  \Fc_{R}\LF \frac{\bar{ R}_{\rm dS}}{4\Mc_{s}^2}\RF - \Fc_{1} +\epsilon\frac{\bar{ R}_{\rm dS}}{8\Mc_{s}^2}\Fc^{(\dagger)}_R\LF \frac{\bar{ R}_{\rm dS}}{4\Mc_{s}^2}\RF \RT  \, .
	\label{TcNL}
\end{equation}
\begin{equation}
	\begin{aligned}
		T_{\rm NL}^1 & = \sum_{i,j,k, i\neq j\neq k} \Lc_{i} \LF \frac{2M^2}{\Mc_{s}^2} \RF  \Lc_{j} \LF \frac{2M^2}{\Mc_{s}^2} \RF  \Lc_{k} \LF \frac{2M^2}{\Mc_{s}^2} \RF \, ,\\ 
		T_{\rm NL}^2 & = \sum_{i,j,k, i\neq j\neq k} \Lc_{i} \LF \frac{2M^2}{\Mc_{s}^2} \RF  \Lc_{j} \LF \frac{2M^2}{\Mc_{s}^2} \RF \Lc_{k} \LF \frac{M^2}{\Mc_{s}^2} + \frac{\bar{ R}_{\rm dS}}{4\Mc_{s}^2}\RF \, ,\\ 
		T_{\rm NL}^3 & = \sum_{i,j,k, i\neq j\neq k} \Lc_{i} \LF \frac{2M^2}{\Mc_{s}^2} \RF  \Lc_{j} \LF \frac{M^2}{\Mc_{s}^2} + \frac{\bar{ R}_{\rm dS}}{4\Mc_{s}^2} \RF \Lc_{k} \LF \frac{M^2}{\Mc_{s}^2} + \frac{\bar{ R}_{\rm dS}}{4\Mc_{s}^2}\RF\, , \\
		T_{\rm NL}^4 & = \sum_{i,j,k, i\neq j\neq k} \Lc_{i} \LF  \frac{M^2}{\Mc_{s}^2} + \frac{\bar{ R}_{\rm dS}}{4\Mc_{s}^2} \RF  \Lc_{j}\LF \frac{M^2}{\Mc_{s}^2} + \frac{\bar{ R}_{\rm dS}}{4\Mc_{s}^2} \RF \Lc_{k} \LF  \frac{M^2}{\Mc_{s}^2}+ \frac{\bar{ R}_{\rm dS}}{4\Mc_{s}^2} \RF\,. 
	\end{aligned}
	\label{tnls}
\end{equation}
Considering the formfactors of the form \eqref{cubicformd} with the conditions imposed in \eqref{entic}, we can easily deduce that
\begin{equation}
	T_{\rm NL}^1 \ll  T_{\rm NL}^2\ll T_{\rm NL}^3\ll T_{\rm NL}^4\,
	\label{ineq}
\end{equation}
in the limit of $M^2\ll \Mc_s^2$ and $\bar{R}_{\rm dS}\gtrsim \Mc_s^2$. Note that since 
	$T_{\rm NL}^1$ does not depend on the ratio $\frac{\bar{R}_{\rm dS}}{\Mc_s^2}$, we found its contributions are far less than the contributions involving $T_{\rm NL}^2-T_{\rm NL}^4$, especially for the operators $\ell_i\LF \square_s \RF$ of the form in \eqref{entchoice}.  

\section{On violation of Maldacena consistency relation in non-local gravity}
\label{app:violmc} 

The goal of this section is to illustrate how the Maldacena consistency relation can be easily violated in non-local gravity. 
%against the established notions of Maldacena consistency relation in the context of single field inflation. 
We provide here an explicit illustrative example from our study carried in this paper. 
It has been argued that the Maldacena consistency relation given by \cite{Maldacena:2002vr,Creminelli:2004yq,Creminelli:2011rh} 
\begin{equation}
	f_{\rm NL}^{\rm sq} = \frac{5}{12}\LF 1-n_s \RF
\end{equation}
cannot be violated in any single field slow-roll inflation with adiabatic vacuum initial conditions. The above relation was intuitively argued to be valid as follows. Since the curvature perturbation is constant on super-Hubble scales, the perturbed metric (say with wave-number $k_1$) in the so-called unitary gauge in which the inflaton perturbations can be put to zero \cite{Creminelli:2004yq,Creminelli:2011rh} is approximated to be \footnote{C.f. Eq. (17) in~\cite{Starobinsky:1982ee} where $t_0(\bf{r})$ is just $\frac{\zeta({\bf r})}{H}$ up to a constant, and |$\zeta$| need not be small. In that paper, the synchronous gauge with additional conditions excluding two gauge modes was used but the resulting form of the metric in the leading approximation for $t\gg t_0$ appeared to be essentially the same as \eqref{effb}.}
\begin{equation}
	ds^2 = -dt^2+e^{2\zeta_{k_1}}a^2(t) dx^2\,, 
	\label{effb}
\end{equation} 
where the part of the scalar perturbations related to lapse $\Nc$ and shift functions $\Nc_i$ in the well-known Arnowitt-Deser-Misner (ADM) formalism \cite{Maldacena:2002vr} are approximated to zero as they depend on the time and spatial derivatives of $\zeta$ which is constant on the super-Hubble scales. Therefore, according to the standard intuitive proof \cite{Creminelli:2004yq,Creminelli:2011rh} that is well supported by EFT of single-field inflation, the 3-point correlation in the squeezed limit $k_1\ll k_2=k_3 = k$ can be decomposed as (using the simple Taylor expansion of 2-point correlation of short wavelength mode in the effective background metric \eqref{effb})
\begin{equation}
	\begin{aligned}
		\lim_{k_1\to 0}\langle  \zeta_{\textbf{k}_1}\zeta_{\textbf{k}_2}\zeta_{\textbf{k}_3}\rangle  & \approx 
		\lim_{k_1\to 0}\langle \zeta_{\textbf{k}_1}\langle\zeta_{\textbf{k}_2}\zeta_{\textbf{k}_3}\rangle \rangle \\ &=
		-\LF 2\pi \RF^3 \delta^{3}\LF \sum_i \textbf{k}_i \RF \LF n_s-1 \RF P_{k_1}P_{k_3}
	\end{aligned} 
	\label{int-der}
\end{equation}
where $P_{k_1},\,P_{k_3}$ are the power spectrum of $k_1$ mode and $k_3$ mode respectively. In deriving the above result, the trick of rescaling the spatial coordinates $x^i\to e^{\zeta_{k_1}}x^i$ is used since $\zeta_{k_1}$ can be treated as a constant on super-Hubble scales that was shown in~\cite{Starobinsky:1982ee,Starobinsky:2001xq,Creminelli:2011rh}. The above result is also concretely established and proved in local scalar-tensor theories.

In the context of non-local gravity we have seen that non-local interactions of curvature perturbations evade the above result (as we can witness from the expression of $f_{\rm NL}^{\rm sq}$ in \eqref{fnlseo}). To illustrate this, we consider a part of the action \eqref{NAID} with local $R+R^2$ terms and the non-local cubic scalar curvature term 
\begin{equation}
	S_{R+R^2+R^3}^{\rm Non-local}  = S_{R+R^2}^{\rm local} +S_{R^3}^{\rm Non-local}
	\label{exact} 
\end{equation}
where $S_{R+R^2}^{\rm local},\, S_{R^3}^{\rm Non-local} $ are defined in \eqref{listact} with condition \eqref{entic}.  

Computing the second order perturbed action of \eqref{exact} around the FLRW background obtained from \eqref{mR2in}, we can deduce that 
\begin{equation}
	\delta^{(2)}S_{R+R^2+R^3}^{\rm Non-local} = 	\delta^{(2)}S_{R+R^2}^{\rm local}\,,
\end{equation}
where $\delta^{(2)}S_{R^3}^{\rm Non-local} \Big\vert_{\bar{	\square}\bar{R}=M^2\bar{R}} =0 $ which is the result derived in \cite{Koshelev:2022olc}. This means that at the second order perturbation level, \eqref{exact} is exactly equivalent to local $R^2$ inflation and we can easily confirm that curvature perturbation is conserved on super-Hubble scales. But if we go to the interaction level, that is when we compute the bi-spectrum, we generate non-trivial contributions from the non-local interactions generated from the non-local cubic in scalar curvature term. These new contributions violate the Maldacena consistency relation as we can see from expression \eqref{fnlseo}.
%in the limit $\Fc_R\LF \square_s \RF\to f_0$. 
Therefore, we can now precisely deduce that \eqref{exact} is a counterexample to the intuitive derivation of the Maldacena consistency relation \eqref{int-der} because in the case of inflation with \eqref{exact} we can see from the second order action that it is single field and slow-roll inflation since the background solutions is \eqref{mR2in}, but still we do violate the Maldacena consistency relation. This happens because the relations \eqref{Commtr} lead to the enhancement of the interaction strength between the long wavelength mode and short wavelength modes beyond this order of slow-roll since $\bar{R}\gtrsim \Mc_s^2$. 

So in summary, the proof \eqref{int-der} can only be valid in local theories, in particular two derivative theories, where the perturbed metric can be approximated as \eqref{effb}. Then one can Taylor expand 3-point correlation in the squeezed limit. In the case of non-local theory, one cannot apply this simple Taylor expansion because vertices come with formfactors of infinite derivative structure which we  localize the action using the on-shell relations \eqref{Commtr}. This alters significantly the interaction picture and we get an exponential enhancement of 3-point vertices due to the structure of formfactors \eqref{formfinal}. However, we are well within the perturbative regime because the non-linear corrections to curvature perturbation in non-local $R^2$-like inflation are still small and fluctuations are indeed very Gaussian as the current data suggest \cite{Akrami:2019izv}.

\bibliographystyle{utphys}
\bibliography{ssa.bib}

\end{document}